\documentclass[12pt,english,onecolumn, draftcls]{IEEEtran}
\usepackage[latin9]{luainputenc}
\usepackage{color}
\usepackage{babel}
\usepackage{verbatim}
\usepackage{mathrsfs}
\usepackage{amsmath}
\usepackage{amsthm}
\usepackage{amssymb}
\usepackage{graphicx}
\usepackage{setspace}
\usepackage[unicode=true,pdfusetitle,
 bookmarks=true,bookmarksnumbered=false,bookmarksopen=false,
 breaklinks=false,pdfborder={0 0 1},backref=false,colorlinks=false]
 {hyperref}

\makeatletter
\theoremstyle{plain}
\newtheorem{thm}{\protect\theoremname}
\theoremstyle{plain}
\newtheorem{lem}[thm]{\protect\lemmaname}

\usepackage{subfigure}
\usepackage{epstopdf}

\makeatother

\providecommand{\lemmaname}{Lemma}
\providecommand{\theoremname}{Theorem}

\begin{document}

\title{Uplink Performance Analysis of Dense Cellular Networks with LoS and
NLoS Transmissions
}

\author{\noindent {\normalsize{}Tian Ding, }\textit{\normalsize{}Student
Member, IEEE}{\normalsize{}, Ming Ding, }\textit{\normalsize{}Member,
IEEE}{\normalsize{}, }\\
{\normalsize{}Guoqiang Mao, }\textit{\normalsize{}Senior Member, IEEE}{\normalsize{},
Zihuai Lin, }\textit{\normalsize{}Senior Member, IEEE}{\normalsize{},
}\\
{\normalsize{}David L$\acute{\textrm{o}}$pez-P$\acute{\textrm{e}}$rez,
}\textit{\normalsize{}Member, IEEE}{\normalsize{}, Albert Zomaya,
}\textit{\normalsize{}Fellow, IEEE}\thanks{Tian Ding is with The University of Technology Sydney, Australia (e-mail:
Tian.Ding@student.uts.edu.au). }\thanks{Ming Ding is with Data61, Sydney, Australia (e-mail: Ming.Ding@nicta.com.au). }\thanks{Guoqiang Mao is with the School of Computing and Communication, The
University of Technology Sydney, Australia. He also holds adjunct
positions at the School of Electronic Information \& Communications,
Huazhong University of Science \& Technology, Wuhan, China, and the
School of Information and Communication Engineering, Beijing University
of Posts and Telecommunications, Beijing, China (e-mail: g.mao@ieee.org).
Guoqiang Mao's research is supported by Australian Research Council
(ARC) Discovery project DP110100538 and Chinese National Science Foundation
project 61428102.}\thanks{Zihuai Lin is with the School of Electrical and Information Engineering,
The University of Sydney, Australia (e-mail: zihuai.lin@sydney.edu.au).}\thanks{David L$\acute{\textrm{o}}$pez-P$\acute{\textrm{e}}$rez is with
Bell Labs, Nokia, Dublin, Ireland (email: dr.david.lopez@ieee.org). }\thanks{Albert Zomaya is with the School of IT, the University of Sydney,
Australia (email: albert.zomaya@sydney.edu.au). }
}

\maketitle
{} 
\begin{abstract}
\textcolor{red}{}%
In this paper, we analyse the coverage probability and the area spectral
efficiency (ASE) for the uplink (UL) of dense small cell networks
(SCNs) considering a practical path loss model incorporating both
line-of-sight (LoS) and non-line-of-sight (NLoS) transmissions. Compared
with the existing work, we adopt the following novel approaches in
our study: (i) we assume a practical user association strategy (UAS)
based on the smallest path loss, or equivalently the strongest received
signal strength; (ii) we model the positions of both base stations
(BSs) and the user equipments (UEs) as two independent Homogeneous
Poisson point processes (HPPPs); and (iii) the correlation of BSs'
and UEs' positions is considered, thus making our analytical results
more accurate. The performance impact of LoS and NLoS transmissions
on the ASE for the UL of dense SCNs is shown to be significant, both
quantitatively and qualitatively, compared with existing work that
does not differentiate LoS and NLoS transmissions. In particular,
existing work predicted that a larger UL power compensation factor
would always result in a better ASE in the practical range of BS density,
i.e., $10^{1}\sim10^{3}\,\textrm{BSs/km}^{2}$. However, our results
show that a smaller UL power compensation factor can greatly boost
the ASE in the UL of dense SCNs, i.e., $10^{2}\sim10^{3}\,\textrm{BSs/km}^{2}$,
while a larger UL power compensation factor is more suitable for sparse
SCNs, i.e., $10^{1}\sim10^{2}\,\textrm{BSs/km}^{2}$.
\end{abstract}


\section{Introduction}

By means of network densification, small cell networks (SCNs) can
achieve a high spatial reuse gain, which further leads to a high network
capacity~\cite{Tutor_smallcell}. Particularly, the orthogonal deployment
of SCNs within the existing macrocell network, i.e., small cells and
macrocells operating on different frequency spectrum (Small Cell Scenario
\#2a defined in~\cite{TR36.872}), is prioritized in the design of
the 4th generation (4G) Long Term Evolution (LTE) networks by the
3rd Generation Partnership Project (3GPP). Furthermore, dense SCNs
are envisaged to be the workhorse for capacity enhancement in the
5th generation (5G) networks due to its large performance gains and
easy deployment~\cite{Tutor_smallcell,Xiaohu_5Gdense,Smallcell_Backhaul}.
Thus, this paper focuses on studying the performance of these orthogonal
deployments of dense SCNs.

In our previous work~\cite{Ming_DL}, we conducted a study on the
downlink (DL) of dense SCNs considering a sophisticated path loss
model that differentiates line-of-sight (LoS) and non-line-of-sight
(NLoS) transmissions. LoS transmission may occur when the distance
between a transmitter and a receiver is small, and NLoS transmission
is more common in office environments and in central business districts.
Moreover, the probability that there exists a LoS path between the
transmitter and the receiver increases as their distance decreases.
It is observed in~\cite{Ming_DL} that the reduction of the distance
between the transmitter and the receiver as the density of small cell
base stations (BSs) increases will cause a transition from NLoS transmission
to LoS transmission, which has a significant impact, both quantitatively
and qualitatively, on the performance of DL dense SCNs. Motivated
by this finding~\cite{Ming_DL}, in this paper, we continue to query
whether such NLoS-to-LoS transitions may significantly affect the
performance of uplink (UL) dense SCNs.

Our work distinguishes from existing work~\cite{Ming_DL,Jeff_UL,UL_Tian_ICC}
on the performance analysis of UL dense SCNs in three major aspects.
First, we assume a user association strategy (UAS) that each UE is
associated with the BS with the smallest path loss to the UE, or equivalently
each UE is associated with the BS that delivers the strongest received
signal strength~\cite{Ming_DL}. Note that in our previous work~\cite{UL_Tian_ICC}
and existing work in the literature~\cite{Jeff_UL} , the authors
assumed that each UE should be associated with the closest BS. Such
assumption is not appropriate for the realistic path loss model with
LoS and NLoS transmissions, because in practice it is possible for
a UE to associate with a BS that is not the closest one but with a
LoS path, instead of the nearest BS with a NLoS path. Second, we assume
that the BSs and the UEs are deployed according to two independent
Homogeneous Poisson point processes (HPPPs){\small{},} which is more
practical and realistic compared with the previous work~\cite{Jeff_UL,UL_Tian_ICC}.
Third, we consider the correlation of BS and UE positions explained
later in the paper, thus making our numerical results more accurate
than the previous work~\cite{Jeff_UL}, which ignored such correlation.
The main contributions of this paper are as follows: 
\begin{itemize}
\item Numerically tractable results are obtained for the UL coverage probability
and the UL area spectral efficiency (ASE) performance using a piecewise
path loss model, incorporating both LoS and NLoS transmissions.
\item Our theoretical analysis of the UL of dense SCNs shows a similar performance
trend that was found for the DL of dense SCNs in our previous work~\cite{Ming_DL},
i.e., when the density of UEs is larger than a threshold, %
the ASE may suffer from a slow growth or even a decrease. Then, the
ASE will grow almost linearly as the UE density increases above another
larger threshold. This finding is in stark contrast with previous
results using a simplistic path loss model that does not differentiate
LoS and NLoS transmissions~\cite{Jeff_UL}. %
\item Our theoretical analysis also indicates that the performance impact
of LoS and NLoS transmissions on the UL of SCNs with \emph{UL power
compensation} is significant both quantitatively and qualitatively
compared with existing work in the literature that does not differentiate
LoS and NLoS transmissions. The details of the UL power compensation
scheme will be introduced in Section~\ref{sec:Network-Model}. In
particular, the previous work~\cite{Jeff_UL} showed that a larger
UL power compensation factor should always deliver a better ASE performance
in the practical range of BS density, i.e., $10^{1}\sim10^{3}\,\textrm{BSs/km}^{2}$.
However, our results show that a smaller UL power compensation factor
can greatly boost the ASE performance in dense SCNs, i.e., $10^{2}\sim10^{3}\,\textrm{BSs/km}^{2}$,
while a larger UL power compensation factor is more suitable for sparse
SCNs, i.e., $10^{1}\sim10^{2}\,\textrm{BSs/km}^{2}$. Our new finding
indicates that it is possible to save UE battery and meanwhile obtain
a high ASE in the UL of dense SCNs in 5G, if the UL power compensation
factor is optimized.
\end{itemize}

The remainder of this paper is structured as follows. Section~\ref{sec:Related-Work}
compares the closest related work to our work. Section~\ref{sec:Network-Model}
describes the system model. Section \ref{sec:Analysis-Proposed} presents
our main analytical results on the UL coverage probability and the
UL ASE. Section~\ref{sec:UL-Analysis} presents the application of
our main analytical results on the UL coverage probability and the
UL ASE in a 3GPP special case, followed by a more efficient computation
method to evaluate the results using the Gauss-Laguerre quadrature.
The numerical results and simulation results are discussed in Section~\ref{sec:Simulaiton-and-Discussion}.
Finally, the conclusions are drawn in Section~\ref{sec:Conclusion}.

\section{Related Work\label{sec:Related-Work}}

In the DL performance analysis of cellular networks based on stochastic
geometry, BS positions are typically modeled as a Homogeneous Poisson
point process (HPPP) on the plane~\cite{Jeff_DL}, and in this case,
the coverage probability can be expressed in a closed-form. Furthermore,
an important and novel capacity model was proposed for HPPP random
cellular networks, where the impact of random interference on the
cooperative communications is analyzed by a closed-form expression~\cite{Xiaohu_Capacity}.
In the UL performance analysis of cellular networks based on stochastic
geometry, UE positions are typically modeled as a HPPP on the plane~\cite{Jeff_UL},
and BS positions are assumed to be uniformly and randomly deployed
in the Voronoi cell of each UE. The difficulty of modeling both BSs
and UEs as a HPPP is that the BS and UE positions are coupled~\cite{Jeff_UL,DynamicTDD},
and the dependence of UE positions is therefore hard to analyse~\cite{Mao_Sensor_Network_Localization,Roadtraffic,Dynamic_Multihop}.
Such dependence occurs because if a UE is associated with a BS that
delivers the strongest received signal (or is closest to the UE),
it implies that there are no other BSs that can be located in positions
that deliver the strongest received signal (or in a closer distance
than the aforementioned BS). To derive tractable and closed-form results,
previous work ignored this dependence and modeled the distance between
a UE and its associated BS as an independent identical distributed
(i.i.d.) random variable. 

In greater detail, in~\cite{Jeff_UL}, the authors assumed that the
UEs are randomly distributed following a HPPP, and exactly one BS
is randomly and uniformly located in each UE's Voronoi cell, i.e.,
each BS associates with its nearest UE. It is also assumed that the
distance between each BS and its serving UE is i.i.d. Rayleigh distributed.
The system model of only deploying UEs as a HPPP~\cite{Jeff_UL}
makes it difficult to conduct network performance analysis for UL
of SCNs. Furthermore, the association strategy that each BS associates
with its nearest UE~\cite{Jeff_UL} is impractical, and the assumption
that all of the BS-UE association distances are i.i.d. Rayleigh distributed~\cite{Jeff_UL}
is unrealistic.

In~\cite{UL_spatial_blocking}, the authors considered UE spatial
blocking, which is referred to as the outage caused by limited number
of usable channels, and derived approximate expressions for the UL
blocking probability and the UL coverage probability. In~\cite{UL_load_balance},
the authors proposed a tractable model to characterize the UL rate
distribution in a $K$-tier heterogeneous cellular networks (HCNs)
considering power control and load balancing. In~\cite{UL_truncated},
the authors considered the maximum power limitation for UEs and obtained
approximate expressions for the UL outage probability and UL spectral
efficiency. However, none of the aforementioned UL related work considered
a realistic path loss model with line-of-sight (LoS) and non-line-of-sight
(NLoS) transmissions. In contrast, in this paper, we consider a sophisticated
path loss model incorporating both LoS and NLoS transmissions to study
their performance impact on dense SCNs and show that LoS and NLoS
transmissions have a significant impact on the performance of UL dense
SCNs.

LoS and NLoS transmissions have been previously investigated in the
DL performance analysis of dense SCNs~\cite{Ming_DL,Jesus_LoS_DL_ICC16}.
One major conclusion of~\cite{Ming_DL} is that the ASE performance
will slowly increase or even decrease in certain BS density regions.
It is interesting to see whether this conclusion holds for UL dense
SCNs. In our previous work on the UL performance analysis of dense
SCNs~\cite{UL_Tian_ICC}, we assume that each UE is associated with
its nearest BS, which may not be a practical assumption when considering
LoS and NLoS transmissions. Compared with~\cite{UL_Tian_ICC}, in
this work we consider a more realistic user association strategy,
in which a UE associates with the BS that has the smallest path loss,
or equivalently delivers the strongest received signal strength. This
user association strategy is more realistic and is particularly important
when considering both LoS and NLoS transmissions that are present
in realistic radio environment, because the closest BS may possibly
have only a NLoS path to the UE and therefore may offer a weaker signal
than a BS that is further away but has a LoS path to the UE.

\section{System Model\label{sec:Network-Model}}

\begin{doublespace}
Different from the assumption that only UEs' deployment follows HPPP
distribution~\cite{Jeff_UL}, in this paper, we assume that both
BSs and UEs are distributed following HPPPs with densities $\lambda$\ $\mathrm{BSs/km^{2}}$
and $\lambda^{\textrm{UE}}$\ $\mathrm{UEs/km^{2}}$, respectively.
Here, we assume that $\lambda^{\textrm{UE}}\gg\lambda$ so that all
the BSs are activated to serve at least one UE. Each UE is assumed
to associate with the BS with the smallest path loss. We focus on
UL and consider a randomly tagged BS, which is denoted as the typical
BS located at the origin. With the assumption of $\lambda^{\textrm{UE}}\gg\lambda$,
on each time-frequency resource block, each BS has one active UE in
its coverage area. The UE associated with the typical BS is denoted
as the typical UE, and the other UEs using the same time-frequency
resource block are denoted as the interfering UEs. The distance from
the typical UE to the typical BS is denoted by $R$, which is a random
variable whose distribution will be analyzed later. Throughout the
paper, we use the upper case letters, e.g., $R$, to denote a random
variable and use the lower case letters, e.g., $r$, to denote specific
instance of the random variable.

The link from the typical UE to the typical BS has a LoS path or a
NLoS path with probability $\mathrm{Pr^{L}}\left(r\right)$ and $\mathrm{1-Pr^{L}}\left(r\right)$,
respectively, where such probability can be computed by the following
piecewise function~\cite{Ming_DL},

\begin{equation}
\mathrm{Pr^{L}}\left(r\right)=\begin{cases}
\mathrm{Pr_{1}^{L}}\left(r\right), & 0<r\leq d_{1}\\
\mathrm{Pr_{2}^{L}}\left(r\right), & d_{1}<r<d_{2}\\
\vdots & \vdots\\
\mathrm{Pr_{\mathit{N}}^{L}}\left(r\right), & r>d_{N-1}
\end{cases}.\label{eq:Pr_of_LoS}
\end{equation}

The distance dependent path loss is expressed as $\zeta\left(r\right)$
with $r$ being the distance, and the path loss gain is $\zeta\left(r\right)^{-1}$,
where the %
path loss of each link is modeled as

\begin{equation}
\zeta\left(r\right)=\begin{cases}
\begin{cases}
A_{1}^{\textrm{L}}r^{\alpha_{1}^{\textrm{L}}}, & \mathrm{LoS\:\mathrm{with\:}probability\:Pr_{1}^{L}\left(\mathit{r}\right)}\\
A_{1}^{\textrm{N}\textrm{L}}r^{\alpha_{1}^{\textrm{NL}}}, & \mathrm{NLoS\:with\:}\mathrm{probability\:\left(1-Pr_{1}^{L}\left(\mathit{r}\right)\right)}
\end{cases}, & 0<r\leq d_{1}\\
\begin{cases}
A_{2}^{\textrm{L}}r^{\alpha_{2}^{\textrm{L}}}, & \mathrm{LoS\:\mathrm{with\:}probability\:Pr_{2}^{L}\left(\mathit{r}\right)}\\
A_{2}^{\textrm{N}\textrm{L}}r^{\alpha_{2}^{\textrm{NL}}}, & \mathrm{NLoS\:with\:}\mathrm{probability\:\left(1-Pr_{2}^{L}\left(\mathit{r}\right)\right)}
\end{cases}, & d_{1}<r<d_{2}\\
\vdots & \vdots\\
\begin{cases}
A_{N}^{\textrm{L}}r^{\alpha_{N}^{\textrm{L}}}, & \mathrm{LoS\:\mathrm{with\:}probability\:Pr_{\mathit{N}}^{L}\left(\mathit{r}\right)}\\
A_{N}^{\textrm{N}\textrm{L}}r^{\alpha_{N}^{\textrm{NL}}}, & \mathrm{NLoS\:with\:}\mathrm{probability\:\left(1-Pr_{\mathit{N}}^{L}\left(\mathit{r}\right)\right)}
\end{cases}, & r>d_{N-1}
\end{cases},\label{eq:pathloss}
\end{equation}

\end{doublespace}

\begin{doublespace}
\noindent where for $n\in\left\{ 1,2,\cdots,N\right\} $, $A_{n}^{\textrm{L}}$
is the path loss of LoS path at a reference distance of $r=1$, $A_{n}^{\textrm{NL}}$
is the path loss of NLoS path at a reference distance of $r=1$, $\alpha_{n}^{\textrm{L}}$
is the path loss exponent of LoS link, and $\alpha_{n}^{\textrm{NL}}$
is the path loss exponent of NLoS link. \textcolor{red}{}%

\end{doublespace}

\textit{\emph{The UL transmission power}} of UE $k$ located at a
distance of $r$ is denoted by $P_{k}$, and is subject to a semi-static
power control (PC) mechanism, i.e., the fractional path loss compensation
(FPC) scheme~\cite{TR36.828}. Based on this FPC scheme, $P_{k}$
is modeled as

\noindent 
\begin{equation}
P_{k}=P_{0}\zeta\left(r\right)^{\epsilon},\label{eq:UL_P_UE}
\end{equation}
where $P_{0}$ is the baseline power on the considered RB at the UE,
$\epsilon\in\left(0,1\right]$ is the FPC factor, and $\zeta\left(r\right)$
is expressed in (\ref{eq:pathloss}). 

\begin{doublespace}
In (\ref{eq:UL_P_UE}), the distance-based fractional power compensation
term $\zeta\left(r\right)^{\epsilon}$ is denoted by $\beta\left(r\right)$
and written as%
{} 

\begin{equation}
\beta\left(r\right)=\zeta\left(r\right)^{\epsilon}.\label{eq:power_compensation}
\end{equation}

Therefore, the received signal power at the typical BS can be written
as

\begin{equation}
\begin{array}{l}
P^{\textrm{sig}}=P_{0}\beta\left(R\right)\zeta\left(R\right)^{-1}g\\
=P_{0}\zeta\left(R\right)^{\left(\epsilon-1\right)}g,
\end{array}\label{eq:signal_power}
\end{equation}

\end{doublespace}

\begin{doublespace}
\noindent where \textit{$g$} denotes the channel gain of the multi-path
fading channel and is an i.i.d. exponential distributed random variable.
Hence,\textit{ $g$} follows an exponential distribution with unit
mean.
\end{doublespace}

\begin{doublespace}
As a result, the SINR at the typical BS of the typical UE can be expressed
as

\begin{equation}
\textrm{SINR}=\frac{P^{\textrm{sig}}}{\sigma^{2}+I_{Z}},\label{eq:SINR_UL}
\end{equation}

\end{doublespace}

\begin{doublespace}
\noindent where $\sigma^{2}$ is the noise power, $Z$ is the set
of interfering UEs, and $I_{Z}$ is the interference given by
\end{doublespace}

\begin{doublespace}
\begin{equation}
I_{Z}=\underset{Z}{\sum}P_{0}\beta\left(R_{z}\right)\zeta\left(D_{z}\right)^{-1}g_{z},\label{eq:interference_power}
\end{equation}

\end{doublespace}

\begin{doublespace}
\noindent where \textit{$g_{z}$} denotes the channel gain of the
multi-path fading channel of interferer $z\in Z$, and is an i.i.d.
exponential distributed random variable, which follows an exponential
distribution with unit mean. The distance of interferer $z\in Z$
to its serving BS is denoted by $R_{z}$, and the distance of interferer
$z\in Z$ to the typical BS is denoted by $D_{z}$. The details of
the distribution of $R_{z}$ and $R$ are given in Section \ref{sec:UL-Analysis}.
Since $D_{z}\gg R_{z}$, $D_{z}$ can be approximated by the distance
from the serving BS of interferer $z$ to the typical BS. %

\end{doublespace}

\section{Analysis Based on the Proposed Path Loss Model\label{sec:Analysis-Proposed}}

\begin{doublespace}
The UL coverage probability for the typical BS can be formulated as

\begin{equation}
P^{\textrm{cov}}\left(\lambda,T\right)=\textrm{Pr}\left[\textrm{SINR}>T\right],\label{eq:definition_of_Pcov}
\end{equation}

\end{doublespace}

\begin{doublespace}
\noindent where $T$ is the SINR threshold.
\end{doublespace}

\begin{doublespace}
The area spectral efficiency (ASE) in $\mathrm{bps/Hz/km^{2}}$ for
a given $\lambda$ can be formulated as~\cite{Ming_DL}

\begin{equation}
A^{\mathrm{ASE}}\left(\lambda,T_{0}\right)=\lambda\int_{T_{0}}^{\infty}\log_{2}\left(1+x\right)f_{X}\left(\lambda,x\right)dx,\label{eq:ASE}
\end{equation}

\end{doublespace}

\begin{doublespace}
\noindent where $T_{0}$ is the minimum working SINR for the considered
SCN, and $f_{X}\left(\lambda,x\right)$ is the PDF of the SINR observed
at the typical BS for a particular value of $\lambda$.
\end{doublespace}

\begin{doublespace}
Based on the definition of $P^{\textrm{cov}}\left(\lambda,T\right)$,
which is the complementary cumulative distribution function (CCDF)
of SINR, $f_{X}\left(\lambda,x\right)$ can be computed as

\begin{equation}
f_{X}\left(\lambda,x\right)=\frac{\partial\left(1-P^{\mathrm{cov}}\left(\lambda,x\right)\right)}{\partial x}.\label{eq:PDF_of_SINR}
\end{equation}

Based on the system model presented in Section~\ref{sec:Network-Model},
we can calculate $P^{\textrm{cov}}\left(\lambda,T\right)$ and present
it in the following theorem.
\end{doublespace}
\begin{thm}
\begin{doublespace}
\label{thm:p_cov_UL}$P^{\textrm{cov}}\left(\lambda,T\right)$ can
be derived as

\begin{equation}
P^{\textrm{cov}}\left(\lambda,T\right)=\overset{N}{\underset{n=1}{\sum}}\left(T_{n}^{\mathrm{L}}+T_{n}^{\mathrm{NL}}\right),\label{eq:Pcov_thm1}
\end{equation}

\end{doublespace}

\begin{doublespace}
\noindent where
\end{doublespace}

\begin{doublespace}
\begin{equation}
\begin{array}{l}
T_{n}^{\mathrm{L}}=\int_{d_{n-1}}^{d_{n}}\mathrm{\textrm{Pr}}\left[\left.\frac{P_{0}g\left(A^{\textrm{L}}r^{\alpha^{\mathrm{\textrm{L}}}}\right)^{\left(\epsilon-1\right)}}{\sigma^{2}+I_{Z}}>T\right|\textrm{LoS}\right]f_{R,n}^{\textrm{L}}\left(r\right)\hspace{-0.1cm}dr,\\
T_{n}^{\mathrm{NL}}=\int_{d_{n-1}}^{d_{n}}\textrm{Pr}\left[\left.\frac{P_{0}g\left(A^{\textrm{NL}}r^{\alpha^{\textrm{NL}}}\right)^{\left(\epsilon-1\right)}}{\sigma^{2}+I_{Z}}>T\right|\textrm{NLoS}\right]f_{R,n}^{\textrm{NL}}\left(r\right)dr,
\end{array}\label{eq:3_parts_of_Pcov}
\end{equation}

\end{doublespace}

\noindent and $d_{0}$ and $d_{N}$ are respectively defined as 0
and $\infty$.

\begin{doublespace}
Moreover, $f_{R,n}^{\textrm{L}}\left(r\right)$ and $f_{R,n}^{\textrm{NL}}\left(r\right)$
can be respectively derived as

\begin{equation}
\begin{array}{l}
f_{R,n}^{\textrm{L}}\left(r\right)\\
=\exp\left(-\int_{0}^{r_{1}}\left(1-\textrm{P}\mathrm{r}_{n}^{\mathrm{L}}\left(u\right)\right)2\pi u\lambda du\right)\exp\left(-\int_{0}^{r}\textrm{P}\mathrm{r}_{n}^{\mathrm{L}}\left(u\right)2\pi u\lambda du\right)\\
\hspace{0.5cm}\times\textrm{P}\mathrm{r}_{n}^{\mathrm{L}}\left(r\right)2\pi r\lambda,~~\left(d_{n-1}<r<d_{n}\right),
\end{array}\label{eq:fRn_L}
\end{equation}

\end{doublespace}

\noindent and

\begin{doublespace}
\begin{equation}
\begin{array}{l}
f_{R,n}^{\textrm{NL}}\left(r\right)\\
=\exp\left(-\int_{0}^{r_{2}}\textrm{P}\mathrm{r}_{n}^{\mathrm{L}}\left(u\right)2\pi u\lambda du\right)\exp\left(-\int_{0}^{r}\left(1-\textrm{P}\mathrm{r}_{n}^{\mathrm{L}}\left(u\right)\right)2\pi u\lambda du\right)\\
\hspace{0.5cm}\times\left(1-\textrm{P}\mathrm{r}_{n}^{\mathrm{L}}\left(r\right)\right)2\pi r\lambda,~~\left(d_{n-1}<r<d_{n}\right),
\end{array}\label{eq:fRn_NL}
\end{equation}

\end{doublespace}

\begin{doublespace}
\noindent where $r_{1}$ and $r_{2}$ are determined respectively
by
\end{doublespace}

\begin{doublespace}
\begin{equation}
r_{1}=\left(A^{\textrm{L}}r^{\alpha^{\textrm{L}}}/A^{\textrm{NL}}\right)^{1/\alpha^{\textrm{NL}}},\label{eq:r1}
\end{equation}

\end{doublespace}

\noindent and

\begin{doublespace}
\begin{equation}
r_{2}=\left(A^{\textrm{NL}}r^{\alpha^{\textrm{NL}}}/A^{\textrm{L}}\right)^{1/\alpha^{\textrm{L}}}.\label{eq:r2}
\end{equation}

\end{doublespace}

Furthermore,\textcolor{black}{{} $\mathrm{\textrm{Pr}}\left[\left.\frac{P_{0}g\left(A^{\textrm{L}}r^{\alpha^{\mathrm{\textrm{L}}}}\right)^{\left(\epsilon-1\right)}}{\sigma^{2}+I_{Z}}>T\right|\textrm{LoS}\right]$
and $\textrm{Pr}\left[\left.\frac{P_{0}g\left(A^{\textrm{NL}}r^{\alpha^{\textrm{NL}}}\right)^{\left(\epsilon-1\right)}}{\sigma^{2}+I_{Z}}>T\right|\textrm{NLoS}\right]$
}are respectively computed by

\begin{equation}
\mathrm{\textrm{Pr}}\left[\left.\frac{P_{0}g\left(A^{\textrm{L}}r^{\alpha^{\mathrm{\textrm{L}}}}\right)^{\left(\epsilon-1\right)}}{\sigma^{2}+I_{Z}}>T\right|\textrm{LoS}\right]=\exp\left(-\frac{T\sigma^{2}}{P_{0}\left(A^{\textrm{L}}r^{\alpha^{\textrm{L}}}\right)^{\left(\epsilon-1\right)}}\right)\mathscr{L}_{I_{Z}}\left(\frac{T}{P_{0}\left(A^{\textrm{L}}r^{\alpha^{\textrm{L}}}\right)^{(\epsilon-1)}}\right),\label{eq:Pcov_LoS}
\end{equation}

\noindent and

\begin{equation}
\textrm{Pr}\left[\left.\frac{P_{0}g\left(A^{\textrm{NL}}r^{\alpha^{\textrm{NL}}}\right)^{\left(\epsilon-1\right)}}{\sigma^{2}+I_{Z}}>T\right|\textrm{NLoS}\right]=\exp\left(-\frac{T\sigma^{2}}{P_{0}\left(A^{\textrm{NL}}r^{\alpha^{\textrm{NL}}}\right)^{\left(\epsilon-1\right)}}\right)\mathscr{L}_{I_{Z}}\left(\frac{T}{P_{0}\left(A^{\textrm{NL}}r^{\alpha^{\textrm{NL}}}\right)^{(\epsilon-1)}}\right),\label{eq:Pcov_NLoS}
\end{equation}

\noindent where $\mathscr{L}_{I_{Z}}\left(s\right)$ is the Laplace
transform of RV $I_{Z}$ evaluated at $s$.
\end{thm}
\begin{IEEEproof}
\begin{doublespace}
See Appendix A.
\end{doublespace}
\end{IEEEproof}
As can be observed from Theorem \ref{thm:p_cov_UL}, the piece-wise
path loss function for LoS transmission, the piece-wise path loss
function for NLoS transmission, and the piece-wise LoS probability
function play active roles in determining the final result of $P^{\textrm{cov}}\left(\lambda,T\right)$.
We will investigate their impacts on network performance in detail
in the following sections. Plugging $P^{\textrm{cov}}\left(\lambda,T\right)$
obtained from (\ref{eq:Pcov_thm1}) into (\ref{eq:PDF_of_SINR}) ,
we can get the result of the ASE using (\ref{eq:ASE}).

\begin{doublespace}

\end{doublespace}
\begin{doublespace}

\section{Study of A 3GPP Special Case\label{sec:UL-Analysis}}
\end{doublespace}

As a special case for Theorem \ref{thm:p_cov_UL}, we consider a path
loss function adopted in the 3GPP as~\cite{TR36.828}

\begin{doublespace}
\begin{equation}
\zeta\left(r\right)=\begin{cases}
A^{\textrm{L}}r^{\alpha^{\textrm{L}}}, & \mathrm{LoS\:\mathrm{with\:}probability\:Pr^{L}\left(\mathit{r}\right)}\\
A^{\textrm{N}\textrm{L}}r^{\alpha^{\textrm{NL}}}, & \mathrm{NLoS\:with\:}\mathrm{probability\:\left(1-Pr^{L}\left(\mathit{r}\right)\right)}
\end{cases},\label{eq:pathloss-1}
\end{equation}

\end{doublespace}

\noindent together with a linear LoS probability function of $\textrm{P\ensuremath{r^{L}}}\left(\mathit{r}\right)$,
defined in the 3GPP as~\cite{SCM_pathloss_model}

\begin{doublespace}
\begin{equation}
\mathrm{Pr^{L}}\left(r\right)=\begin{cases}
1-\frac{r}{d_{1}}, & 0<r\leq d_{1}\\
0, & r>d_{1}
\end{cases},\label{eq:linear_Pr_of_LoS-1}
\end{equation}

\end{doublespace}

\begin{doublespace}
\noindent where $d_{1}$ is the cut-off distance of the LoS link.
\end{doublespace}

For the 3GPP special case, according to Theorem \ref{thm:p_cov_UL},
$P^{\textrm{cov}}\left(\lambda,\gamma\right)$ can then be computed
by

\begin{equation}
P^{\textrm{cov}}\left(\lambda,T\right)=\overset{2}{\underset{n=1}{\sum}}\left(T_{n}^{\mathrm{L}}+T_{n}^{\mathrm{NL}}\right).\label{eq:Pcov_case1}
\end{equation}

In the following subsections, we will investigate the results of $T_{1}^{\mathrm{L}}$,
$T_{1}^{\mathrm{NL}}$, $T_{2}^{\mathrm{L}}$, and $T_{2}^{\mathrm{NL}}$,
respectively.
\begin{doublespace}

\subsection{The Result of $T_{1}^{\mathrm{L}}$}
\end{doublespace}

\begin{doublespace}

Regarding the result of $T_{1}^{\mathrm{L}}$, which is the coverage
probability when the typical UE is associated with the typical BS
with a LoS link of distance less than $d_{1}$, we present Lemma~\ref{lem:T1L}
in the following.
\end{doublespace}
\begin{lem}
\begin{doublespace}
\label{lem:T1L}When the typical UE is associated with a LoS BS of
a distance less than $d_{1}$, the coverage probability $T_{1}^{\mathrm{L}}$
can be computed by

\begin{equation}
T_{1}^{\mathrm{L}}\hspace{-0.1cm}=\hspace{-0.1cm}\int_{0}^{d_{1}}\hspace{-0.1cm}\hspace{-0.1cm}e^{-\frac{T\sigma^{2}}{P_{0}\left(A^{\textrm{L}}r^{\alpha^{\textrm{L}}}\right)^{\left(\epsilon-1\right)}}}\mathscr{L}_{I_{Z}}\hspace{-0.1cm}\left(\hspace{-0.1cm}\frac{T}{P_{0}\left(A^{\textrm{L}}r^{\alpha^{\textrm{L}}}\right)^{\left(\epsilon-1\right)}}\hspace{-0.1cm}\right)f_{R,1}^{\textrm{L}}\hspace{-0.1cm}\left(r\right)dr,\label{eq:T1L}
\end{equation}

\end{doublespace}

\begin{doublespace}
\noindent where
\end{doublespace}

\noindent 
\begin{equation}
f_{R,1}^{\textrm{L}}\left(r\right)=\exp\left(-\pi\lambda r^{2}+2\pi\lambda\left(\frac{r^{3}}{3d_{1}}-\frac{r_{1}^{3}}{3d_{1}}\right)\right)\left(1-\frac{r}{d_{1}}\right)2\pi r\lambda,\label{eq:fR1_L-1}
\end{equation}
and the Laplace transform $\mathscr{L}_{I_{Z}}\left(s\right)$ is
expressed as

\noindent 
\begin{eqnarray}
\mathscr{L}_{I_{Z}}\left(s\right) & = & \exp\left\{ \hspace{-0.1cm}-2\pi\lambda\int_{r}^{d_{1}}\hspace{-0.1cm}\left(\hspace{-0.1cm}1\hspace{-0.1cm}-\hspace{-0.1cm}\frac{x}{d_{1}}\hspace{-0.1cm}\right)\int_{0}^{\infty}\left(\left.\frac{1}{1+s^{-1}P_{0}^{-1}\beta\left(u\right)^{-1}\zeta\left(x\right)}f_{R_{z}}^{\mathrm{1L}}\left(u\right)du\right|\textrm{LoS}\right)xdx\hspace{-0.1cm}\right\} \nonumber \\
 &  & \hspace{-0.1cm}\times\hspace{-0.1cm}\exp\left\{ \hspace{-0.1cm}-2\pi\lambda\int_{r_{1}}^{d_{1}}\hspace{-0.1cm}\left(\frac{x}{d_{1}}\right)\int_{0}^{\infty}\left(\left.\frac{1}{1+s^{-1}P_{0}^{-1}\beta\left(u\right)^{-1}\zeta\left(x\right)}f_{R_{z}}^{\mathrm{1NL}}\left(u\right)du\right|\textrm{NLoS}\right)xdx\hspace{-0.1cm}\right\} \nonumber \\
 &  & \hspace{-0.1cm}\times\hspace{-0.1cm}\exp\left\{ \hspace{-0.1cm}-2\pi\lambda\int_{d_{1}}^{\infty}\int_{0}^{\infty}\left(\left.\frac{1}{1+s^{-1}P_{0}^{-1}\beta\left(u\right)^{-1}\zeta\left(x\right)}f_{R_{z}}^{\mathrm{2NL}}\left(u\right)du\right|\textrm{NLoS}\right)xdx\hspace{-0.1cm}\right\} .\label{eq:Laplace_transform}
\end{eqnarray}

\begin{doublespace}
According to the HPPP system model, the distribution of $R_{z}$ is
the same as $R$, but bounded by $x$. The PDF of $R_{z}$ can be
written as
\end{doublespace}

\begin{equation}
f_{R_{z}}\left(u\right)=\begin{cases}
f_{R_{z},1}^{\textrm{L}}\left(u\right), & \textrm{LoS, }0<u\leq x\\
f_{R_{z},1}^{1\textrm{NL}}\left(u\right), & \textrm{NLoS, }0<u\leq x_{1}\\
f_{R_{z},1}^{2\textrm{NL}}\left(u\right), & \textrm{NLoS, }y_{1}<u\leq d_{1}\\
f_{R_{z},2}^{\textrm{NL}}\left(u\right), & \textrm{NLoS, }d_{1}<u\leq x
\end{cases},\label{eq:PDF_Rz}
\end{equation}

\noindent where

\noindent 
\begin{equation}
f_{R_{z},1}^{\textrm{L}}\left(u\right)=\exp\left(-\pi\lambda u^{2}+2\pi\lambda\left(\frac{u^{3}}{3d_{1}}-\frac{u_{1}^{3}}{3d_{1}}\right)\right)\left(1-\frac{u}{d_{1}}\right)2\pi u\lambda,\label{eq:PDF_Rz_1L_L}
\end{equation}

\begin{doublespace}
\begin{equation}
f_{R_{z},1}^{1\textrm{NL}}\left(u\right)=\exp\left(-\pi\lambda u_{2}^{2}+2\pi\lambda\left(\frac{u_{2}^{3}}{3d_{1}}-\frac{u^{3}}{3d_{1}}\right)\right)\left(\frac{u}{d_{1}}\right)2\pi u\lambda,\label{eq:PDF_Rz_1L_NL}
\end{equation}

\begin{equation}
f_{R_{z},1}^{2\textrm{NL}}\left(u\right)=\exp\left(2\pi\lambda\left(-\frac{d_{1}^{2}}{6}-\frac{u^{3}}{3d_{1}}\right)\right)\left(\frac{u}{d_{1}}\right)2\pi u\lambda,\label{eq:PDF_Rz_1NL_2NL}
\end{equation}

\end{doublespace}

\noindent and

\noindent 
\begin{equation}
f_{R_{z},2}^{\textrm{NL}}\left(u\right)=\exp\left(-\pi\lambda u^{2}\right)2\pi u\lambda,\label{eq:PDF_Rz_2_NL}
\end{equation}

\begin{doublespace}
\noindent where
\end{doublespace}

\begin{doublespace}
\begin{equation}
u_{1}=\left(A^{\textrm{L}}u^{\alpha^{\textrm{L}}}/A^{\textrm{NL}}\right)^{1/\alpha^{\textrm{NL}}},\label{eq:u_1}
\end{equation}

\end{doublespace}

\noindent and

\begin{doublespace}
\begin{equation}
u_{2}=\left(A^{\textrm{NL}}u^{\alpha^{\textrm{NL}}}/A^{\textrm{L}}\right)^{1/\alpha^{\textrm{L}}},\label{eq:u_2}
\end{equation}

\end{doublespace}

Specifically, when the interference comes from a LoS path, $f_{R_{z}}^{\mathrm{1L}}\left(u\right)$
can be derived as

\begin{doublespace}
\begin{equation}
f_{R_{z}}^{\mathrm{1L}}\left(u\right)=\begin{cases}
f_{R_{z},1}^{\textrm{L}}\left(u\right), & \textrm{LoS, }0<u\leq x\\
f_{R_{z},1}^{1\textrm{NL}}\left(u\right), & \textrm{NLoS, }0<u\leq x_{1}
\end{cases},\label{eq:PDF_Rz_1L}
\end{equation}

\end{doublespace}

\begin{doublespace}
\noindent where
\end{doublespace}

\begin{doublespace}
\begin{equation}
x_{1}=\left(A^{\textrm{L}}x^{\alpha^{\textrm{L}}}/A^{\textrm{NL}}\right)^{1/\alpha^{\textrm{NL}}}.\label{eq:x_1}
\end{equation}

Conditioned on $x\leq d_{1}$, when the interference path is NLoS,
$f_{R_{z}}^{\mathrm{1NL}}\left(u\right)$ can be derived as

\begin{equation}
f_{R_{z}}^{\mathrm{1NL}}\left(u\right)=\begin{cases}
\begin{cases}
f_{R,1}^{\textrm{L}}\left(u\right),\textrm{ LoS, } & 0<u\leq x_{2}\\
f_{R,1}^{1\textrm{NL}}\left(u\right),\textrm{ NLoS, } & 0<u\leq x
\end{cases}, & r_{1}<x\leq y_{1}\\
\begin{cases}
f_{R,1}^{\textrm{L}}\left(u\right),\textrm{ LoS, } & 0<u\leq d\\
f_{R,1}^{1\textrm{NL}}\left(u\right),\textrm{ NLoS, } & 0<u\leq y_{1}\\
f_{R,1}^{2\textrm{NL}}\left(u\right),\textrm{ NLoS, } & y_{1}<u\leq x
\end{cases}, & y_{1}<x\leq d_{1}
\end{cases},\label{eq:PDF_Rz_1NL}
\end{equation}

\end{doublespace}

\begin{doublespace}
\noindent where
\end{doublespace}

\begin{doublespace}
\begin{equation}
y_{1}=\left(A^{\textrm{L}}d_{1}^{\alpha^{\textrm{L}}}/A^{\textrm{NL}}\right)^{1/\alpha^{\textrm{NL}}},\label{eq:y1-1}
\end{equation}

\end{doublespace}

\begin{doublespace}
\noindent and
\end{doublespace}

\begin{doublespace}
\begin{equation}
x_{2}=\left(A^{\textrm{NL}}x^{\alpha^{\textrm{NL}}}/A^{\textrm{L}}\right)^{1/\alpha^{\textrm{L}}}.\label{eq:x_2}
\end{equation}

Conditioned on $x>d_{1}$, when the interference path is NLoS, $f_{R_{z}}^{\mathrm{2NL}}\left(u\right)$
can be derived as

\begin{equation}
f_{R_{z}}^{\mathrm{2NL}}\left(u\right)=\begin{cases}
f_{R_{z},1}^{\textrm{L}}\left(u\right), & \textrm{LoS, }0<u\leq d_{1}\\
f_{R_{z},1}^{1\textrm{NL}}\left(u\right), & \textrm{NLoS, }0<u\leq y_{1}\\
f_{R_{z},}^{2\textrm{NL}}\left(u\right), & \textrm{NLoS, }y_{1}<u\leq d_{1}\\
f_{R_{z},2}^{\textrm{NL}}\left(u\right), & \textrm{NLoS, }d_{1}<u\leq x
\end{cases}.\label{eq:PDF_Rz_2NL}
\end{equation}
\end{doublespace}
\end{lem}
\begin{IEEEproof}
See Appendix B.
\end{IEEEproof}
\begin{doublespace}

\end{doublespace}
\begin{doublespace}

\subsection{The Result of $T_{1}^{\textrm{NL}}$}
\end{doublespace}

\begin{doublespace}
Regarding the result of $T_{1}^{\textrm{NL}}$, which is the coverage
probability when the typical UE is associated with the typical BS
with a NLoS link of distance less than $d_{1}$, we propose Lemma~3
in the following. 
\end{doublespace}
\begin{lem}
\begin{doublespace}
$T_{1}^{\textrm{NL}}$ can be derived as

\begin{equation}
T_{1}^{\textrm{NL}}\hspace{-0.1cm}=\hspace{-0.1cm}\int_{0}^{d_{1}}\hspace{-0.1cm}\hspace{-0.1cm}e^{-\frac{T\sigma^{2}}{P_{0}r^{\alpha^{\textrm{NL}}\left(\epsilon-1\right)\hspace{-0.1cm}}}}\mathscr{L}_{I_{Z}}\hspace{-0.1cm}\left(\hspace{-0.1cm}\frac{T}{P_{0}r^{\alpha^{\textrm{NL}}\left(\epsilon-1\right)}}\hspace{-0.1cm}\right)f_{R,1}^{\textrm{NL}}\left(r\right)dr,\label{eq:T1NL}
\end{equation}

\end{doublespace}

\begin{doublespace}
\noindent where
\end{doublespace}

\noindent 
\begin{equation}
f_{R,1}^{\textrm{NL}}\left(r\right)=\begin{cases}
\exp\left(-\pi\lambda r_{2}^{2}+2\pi\lambda\left(\frac{r_{2}^{3}}{3d_{1}}-\frac{r^{3}}{3d_{1}}\right)\right)\left(\frac{r}{d_{1}}\right)2\pi r\lambda, & 0<r\leq y_{1}\\
\exp\left(-\frac{\pi\lambda d_{1}^{2}}{3}-\frac{2\pi\lambda r^{3}}{3d_{1}}\right)\left(\frac{r}{d_{1}}\right)2\pi r\lambda, & y_{1}<r\leq d_{1}
\end{cases},\label{eq:fR1_NL-1}
\end{equation}
and the Laplace transform $\mathscr{L}_{I_{Z}}\left(s\right)$ for
$0<r\leq y_{1}$ and $y_{1}<r\leq d_{1}$ are respectively expressed
as

\begin{doublespace}
\begin{equation}
\begin{array}{l}
\mathscr{L}_{I_{Z}}\left(s\right)\\
=\exp\left(-2\pi\lambda\int_{r_{2}}^{d_{1}}\left(\hspace{-0.1cm}1\hspace{-0.1cm}-\hspace{-0.1cm}\frac{x}{d_{1}}\hspace{-0.1cm}\right)\int_{0}^{\infty}\left(\left.\frac{1}{1+s^{-1}P_{0}^{-1}\beta\left(u\right)^{-1}\zeta\left(x\right)}f_{R_{z}}^{1\mathrm{L}}\left(u\right)du\right|\textrm{LoS}\right)xdx\right)\\
\times\exp\left(-2\pi\lambda\int_{r}^{d_{1}}\left(\hspace{-0.1cm}\hspace{-0.1cm}\frac{x}{d_{1}}\hspace{-0.1cm}\right)\int_{0}^{\infty}\left(\left.\frac{1}{1+s^{-1}P_{0}^{-1}\beta\left(u\right)^{-1}\zeta\left(x\right)}f_{R_{z}}^{1\mathrm{NL}}\left(u\right)du\right|\textrm{NLoS}\right)xdx\right)\\
\times\exp\left(-2\pi\lambda\int_{d_{1}}^{\infty}\int_{0}^{\infty}\left(\left.\frac{1}{1+s^{-1}P_{0}^{-1}\beta\left(u\right)^{-1}\zeta\left(x\right)}f_{R_{z}}^{\mathrm{2NL}}\left(u\right)du\right|\textrm{NLoS}\right)xdx\right)
\end{array}\label{eq:Laplace_1NL_first_case}
\end{equation}

\end{doublespace}

\begin{doublespace}
\noindent and
\end{doublespace}

\begin{doublespace}
\begin{equation}
\begin{array}{l}
\mathscr{L}_{I_{Z}}\left(s\right)\\
=\exp\left(-2\pi\lambda\int_{r}^{d_{1}}\frac{x}{d_{1}}\int_{0}^{\infty}\left(\left.\frac{1}{1+s^{-1}P_{0}^{-1}\beta\left(u\right)^{-1}\zeta\left(x\right)}f_{R_{z}}^{\mathrm{1NL}}\left(u\right)du\right|\textrm{NLoS}\right)xdx\right)\\
\times\exp\left(-2\pi\lambda\int_{d_{1}}^{\infty}\int_{0}^{\infty}\left(\left.\frac{1}{1+s^{-1}P_{0}^{-1}\beta\left(u\right)^{-1}\zeta\left(x\right)}f_{R_{z}}^{\mathrm{2NL}}\left(u\right)du\right|\textrm{NLoS}\right)xdx\right),
\end{array}\label{eq:Laplace_1NL_second_case}
\end{equation}

\end{doublespace}

\begin{doublespace}
\noindent where $r_{2}=\left(A^{\textrm{L}}r^{\alpha^{\textrm{NL}}}/A^{\textrm{NL}}\right)^{1/\alpha^{\textrm{L}}}$. 
\end{doublespace}
\end{lem}
\begin{IEEEproof}
The proof is very similar to that in Appendix B. Thus it is omitted
for brevity.
\end{IEEEproof}

\subsection{The Result of $T_{2}^{\textrm{L}}$}

\begin{doublespace}
The result of $T_{2}^{\textrm{L}}$ is the coverage probability when
the typical UE is associated with the typical BS with a LoS link of
distance larger than $d_{1}$. From Theorem \ref{thm:p_cov_UL}, $T_{2}^{\textrm{L}}$
can be derived as
\end{doublespace}

\begin{equation}
T_{2}^{\mathrm{L}}=\int_{d_{1}}^{\infty}\mathrm{\textrm{Pr}}\left[\left.\frac{P_{0}g\left(A^{\textrm{L}}r^{\alpha^{\mathrm{\textrm{L}}}}\right)^{\left(\epsilon-1\right)}}{\sigma^{2}+I_{Z}}>T\right|\textrm{LoS}\right]f_{R,2}^{\textrm{L}}\left(r\right)\hspace{-0.1cm}dr.\label{eq:T2L}
\end{equation}

According to Theorem \ref{thm:p_cov_UL} and (\ref{eq:linear_Pr_of_LoS-1}),
$f_{R,2}^{\textrm{L}}\left(r\right)$ can be calculated by

\begin{equation}
\begin{array}{l}
f_{R,2}^{\textrm{L}}\left(r\right)\\
=\exp\left(-\int_{0}^{r_{1}}\left(1-\textrm{P}\textrm{\ensuremath{r^{L}}}\left(u\right)\right)2\pi u\lambda du\right)\exp\left(-\int_{0}^{r}\textrm{P}\textrm{\ensuremath{r^{L}}}\left(u\right)2\pi u\lambda du\right)\times0\times2\pi r\lambda\\
=0,~~\left(r>d_{1}\right).
\end{array}\label{eq:fR2_L}
\end{equation}

Plugging (\ref{eq:fR2_L}) into (\ref{eq:T2L}), yields

\begin{equation}
T_{2}^{\mathrm{L}}=0.\label{eq:T2L_result}
\end{equation}

\begin{doublespace}

\subsection{The Result of $T_{2}^{\textrm{NL}}$}
\end{doublespace}

\begin{doublespace}
Regarding the result of $T_{2}^{\textrm{NL}}$, which is the coverage
probability when the typical UE is associated with the typical BS
with a NLoS link of distance larger than $d_{1}$, we propose Lemma~4
in the following. 
\end{doublespace}
\begin{lem}
\begin{doublespace}
$T_{2}^{\textrm{NL}}$ can be derived as

\begin{equation}
T_{2}^{\textrm{NL}}\hspace{-0.1cm}=\hspace{-0.1cm}\int_{d_{1}}^{\infty}\hspace{-0.1cm}\hspace{-0.1cm}e^{-\frac{T\sigma^{2}}{P_{0}r^{\alpha^{\textrm{NL}}\left(\epsilon-1\right)\hspace{-0.1cm}}}}\mathscr{L}_{I_{Z}}\hspace{-0.1cm}\left(\hspace{-0.1cm}\frac{T}{P_{0}r^{\alpha^{\textrm{NL}}\left(\epsilon-1\right)}}\hspace{-0.1cm}\right)f_{R,2}^{\textrm{NL}}\hspace{-0.1cm}\left(r\right)dr,\label{eq:T2NL}
\end{equation}

\end{doublespace}

\begin{doublespace}
\noindent where
\end{doublespace}

\noindent 
\begin{equation}
f_{R,2}^{\textrm{NL}}\left(r\right)=\exp\left(-\pi\lambda r^{2}\right)2\pi r\lambda,\label{eq:fR2_NL-1}
\end{equation}
and the Laplace transform $\mathscr{L}_{I_{Z}}\left(s\right)$ is
expressed as

\begin{doublespace}
\begin{equation}
\mathscr{L}_{I_{Z}}\left(s\right)=\exp\left(-2\pi\lambda\int_{r}^{\infty}\int_{0}^{\infty}\left(\left.\frac{1}{1+s^{-1}P_{0}^{-1}\beta\left(u\right)^{-1}\zeta\left(x\right)}f_{R_{z}}^{2\textrm{NL}}\left(u\right)du\right|\textrm{NLoS}\right)xdx\right).\label{eq:Laplace_T2NL}
\end{equation}
\end{doublespace}
\end{lem}
\begin{IEEEproof}
The proof is very similar to that in Appendix B. Thus it is omitted
for brevity..
\end{IEEEproof}
\begin{doublespace}

\subsection{Evaluation Using the Gauss-Laguerre Quadrature}
\end{doublespace}

\begin{doublespace}

\end{doublespace}

To improve the tractability of the derived results, we propose to
approximate the infinite integral of outer-most integrals in (\ref{eq:T2NL})
by the Gauss-Laguerre quadrature~\cite{GL_quadrature_reference},
expressed as

\begin{equation}
\int_{0}^{\infty}f\left(u\right)e^{-u}du\approx\overset{n}{\underset{i=1}{\sum}}\omega_{i}f\left(u_{i}\right),\label{eq:Gauss_Laguerre_Quadrature}
\end{equation}

\noindent where $n$ is the degree of Laguerre polynomial, and $u_{i}$
and $\omega_{i}$ are the \textit{i}-th abscissas and weight of the
quadrature. For practical use, $n$ should be set to a value above
10 to ensure good numerical accuracy~\cite{GL_quadrature_reference}. 

To utilize the Gauss-Laguerre quadrature, the outer-most integral
in (\ref{eq:T2NL}) is rewritten by using the change of variable $\tilde{r}=\pi\lambda r^{2}$.
To evaluate (\ref{eq:T2NL}) by means of the Gauss-Laguerre quadrature,
we propose Lemma~\ref{lem:G-L quadrature} in the following. 
\begin{lem}
\label{lem:G-L quadrature}By using the Gauss-Laguerre quadrature
as shown in (\ref{eq:Gauss_Laguerre_Quadrature}), (\ref{eq:T2NL})
can be approximated and simplified as

\noindent 
\begin{eqnarray}
T_{2}^{\textrm{NL}} & \approx & \overset{n}{\underset{i=1}{\sum}}\omega_{i}\exp\left(-\frac{T\sigma^{2}}{P_{0}\left(\sqrt{\left[u_{i}+\pi\lambda\left(d_{1}\right)^{2}\right]\left(\pi\lambda\right)^{-1}}\right)^{\alpha^{\textrm{NL}}\left(\epsilon-1\right)\hspace{-0.1cm}}}-\pi\lambda\left(d_{1}\right)^{2}\right)\nonumber \\
 &  & \times\hspace{-0.1cm}\mathscr{L}_{I_{Z}}\hspace{-0.1cm}\left(\hspace{-0.1cm}\frac{T}{P_{0}\sqrt{\left[u_{i}+\pi\lambda\left(d_{1}\right)^{2}\right]\left(\pi\lambda\right)^{-1}}^{\alpha^{\textrm{NL}}\left(\epsilon-1\right)}}\hspace{-0.1cm}\right).\label{eq:GL_of_T2NL}
\end{eqnarray}
\end{lem}
\begin{IEEEproof}
See Appendix C.
\end{IEEEproof}

Thanks to Lemma~\ref{lem:G-L quadrature}, the 3-fold integral computation
in (\ref{eq:T2NL}) can now be simplified as a 2-fold integral computation,
which improves the tractability of our results. 
\begin{doublespace}

\section{Simulation and Discussion\label{sec:Simulaiton-and-Discussion}}
\end{doublespace}

In this section, we present numerical and simulation results to establish
the accuracy of our analysis and further study the performance of
the UL of dense SCNs. We adopt the following parameters according
to the 3GPP recommendations~\cite{TR36.828,Mao_Pathloss}: $d_{1}$
= 0.3 km, $\alpha^{\mathrm{L}}$ = 2.09, $\alpha^{\mathrm{NL}}$ =
3.75, $P_{0}$ = -76 dBm, $\sigma^{2}$ = -99 dBm (with a noise figure
of 5\ dB at each BS). We first consider a sparse network in subsection
\ref{subsec:Simulation}, and then we analyze a dense network in the
subsections \ref{subsec:Coverage_vs_density} and \ref{subsec:ASE_vs_density}.

\subsection{Validation of the Analytical Results of $P^{\mathrm{cov}}\left(\lambda,T\right)$\label{subsec:Simulation}}

\begin{figure}
\begin{centering}
\includegraphics[width=12cm]{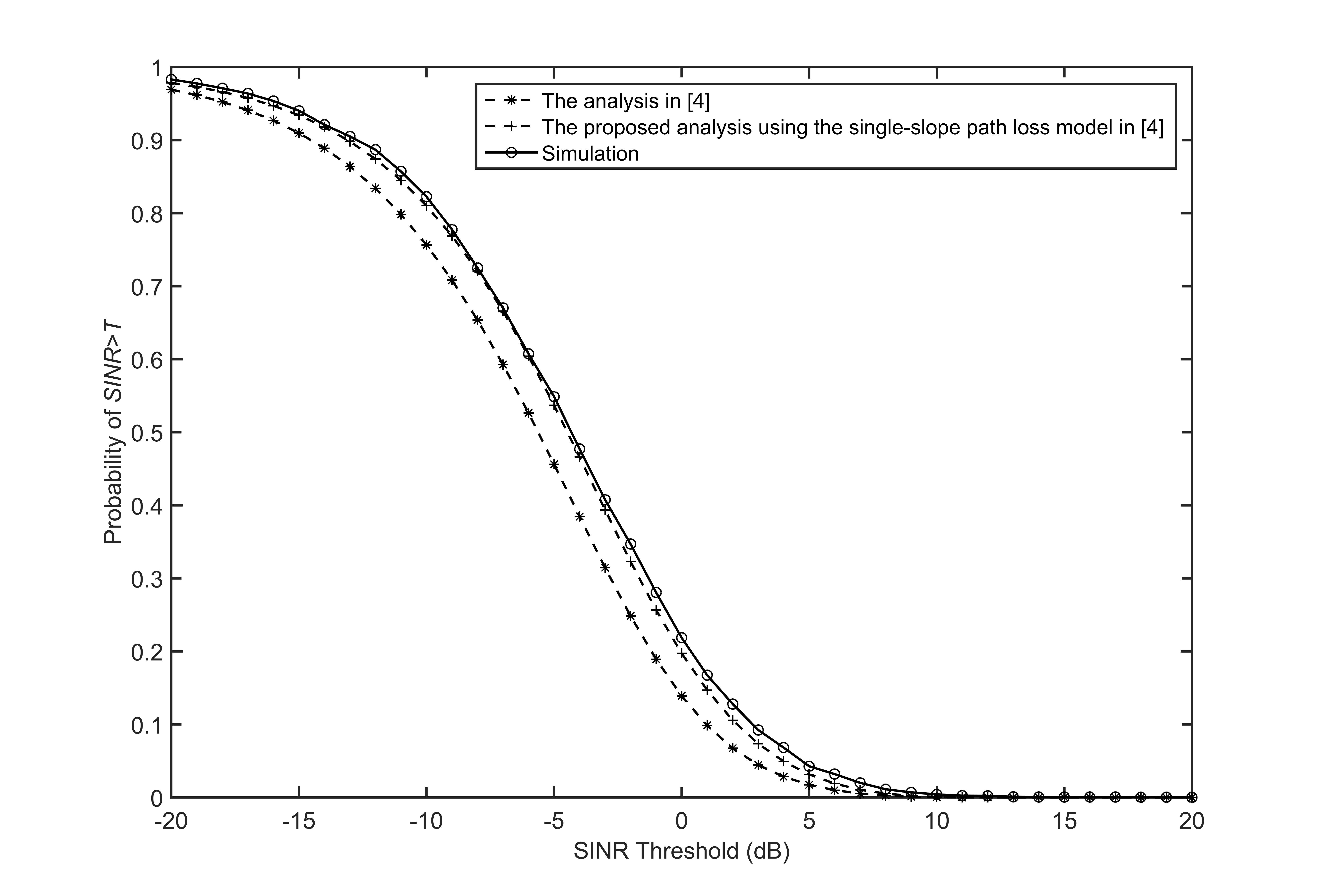}\renewcommand{\figurename}{Fig.}\caption{\label{fig:Simulation_validation_Jeff}The coverage probability $P^{\mathrm{cov}}\left(\lambda,T\right)$
vs. the SINR threshold in~\cite{Jeff_UL} with $\lambda=10\,\mathrm{BSs}/\mathrm{km^{2}}$,
$\alpha=3.75$, and $\epsilon=0.7$.}
\par\end{centering}
\centering{}
\end{figure}

For comparison, we first compute analytical results using a single-slope
path loss model that does not differentiate LoS and NLoS transmissions~\cite{Jeff_UL}.
Note that in~\cite{Jeff_UL}, only one path loss exponent is defined
and denoted by $\alpha$, the value of which is $\alpha=\alpha^{\mathrm{NL}}=3.75$.
The results of $P^{\mathrm{cov}}\left(\lambda,T\right)$ in a sparse
network scenario with $\lambda=10\,\mathrm{BSs}/\mathrm{km^{2}}$,
$\alpha=3.75$, and $\epsilon=0.7$ are plotted in Fig.~\ref{fig:Simulation_validation_Jeff}. 

\textcolor{black}{In the case of the single-slope path loss model}~\textcolor{black}{{[}4{]},
}as can be observed from Fig.~\ref{fig:Simulation_validation_Jeff},
our analytical result is much more accurate than that in~\cite{Jeff_UL}
because our system model assumptions are more reasonable than those
in~\cite{Jeff_UL}: first, the distributions of BSs and UEs are modeled
as two independent HPPPs, instead of the assumption that only UEs
are distributed according to a HPPP~\cite{Jeff_UL}; second, the
dependence of BS and UE positions are discussed, instead of being
ignored~\cite{Jeff_UL}.

\begin{figure}
\centering{}\includegraphics[width=12cm]{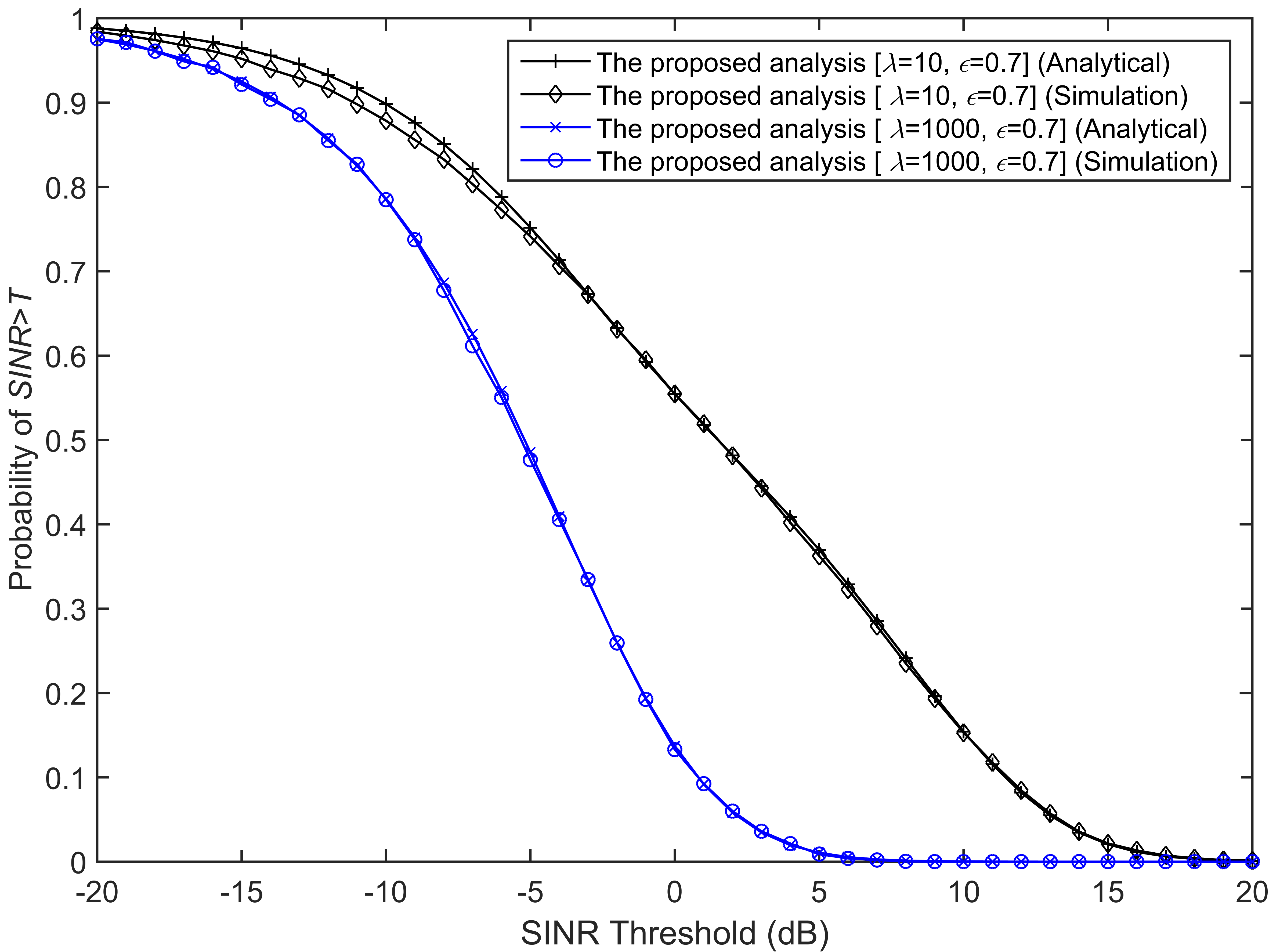}\renewcommand{\figurename}{Fig.}\caption{\label{fig:Simulation validation}The coverage probability $P^{\mathrm{cov}}\left(\lambda,T\right)$
vs. the SINR threshold with $\lambda=10\,\mathrm{BSs}/\mathrm{km^{2}}$
and $\lambda=10^{3}\,\mathrm{BSs}/\mathrm{km^{2}}$.}
\end{figure}

\textcolor{black}{In the case of the 3GPP path loss model}~\cite{TR36.828}\textcolor{black}{,
}the results of $P^{\mathrm{cov}}\left(\lambda,T\right)$ in a sparse
network scenario with $\lambda=10\,\mathrm{BSs}/\mathrm{km^{2}}$
and in a dense network scenario with $\lambda=10^{3}\,\mathrm{BSs}/\mathrm{km^{2}}$
are plotted in Fig.~\ref{fig:Simulation validation}. As can be observed
from Fig.~\ref{fig:Simulation validation}, our analytical results
match the simulation results very well, and thus we will only use
analytical results of $P^{\mathrm{cov}}\left(\lambda,T\right)$ in
our discussion hereafter.

\begin{doublespace}

\end{doublespace}

As can be seen from Fig.~\ref{fig:Simulation validation}, for the
case of $\lambda=10\,\mathrm{BSs}/\mathrm{km^{2}}$, when the SINR
threshold is small (e.g., $T<-4\textrm{dB}$), the analytical result
of the coverage probability is larger than the simulation result.
This is because in our analysis, the approximation of replacing the
location of UE by that of its serving BS, may exclude the cases of
strong interfering UEs located at the proximity of the typical BS,
thus underestimating the total interference, and overestimating the
coverage probability. However, as the SINR threshold increases (e.g.,
$T>-4\textrm{dB}$), the impact of the overestimation of the coverage
probability will decrease, and our analytical result matches the simulation
result well.

Another interesting finding as can be observed from Fig.~\ref{fig:Simulation validation}
is that the analytical result with a larger BS density is more accurate
than that with a smaller BS density. This is because in denser networks,
the distance between a UE and its serving BS is smaller, and the approximation
of replacing the location of a UE by that of its serving BS has less
impact on the estimation of the total interference, thus making the
analytical result more accurate.

\subsection{The Results of $P^{\mathrm{cov}}\left(\lambda,T\right)$ vs. $\lambda$\label{subsec:Coverage_vs_density}}

\begin{figure}
\centering{}\includegraphics[width=12cm]{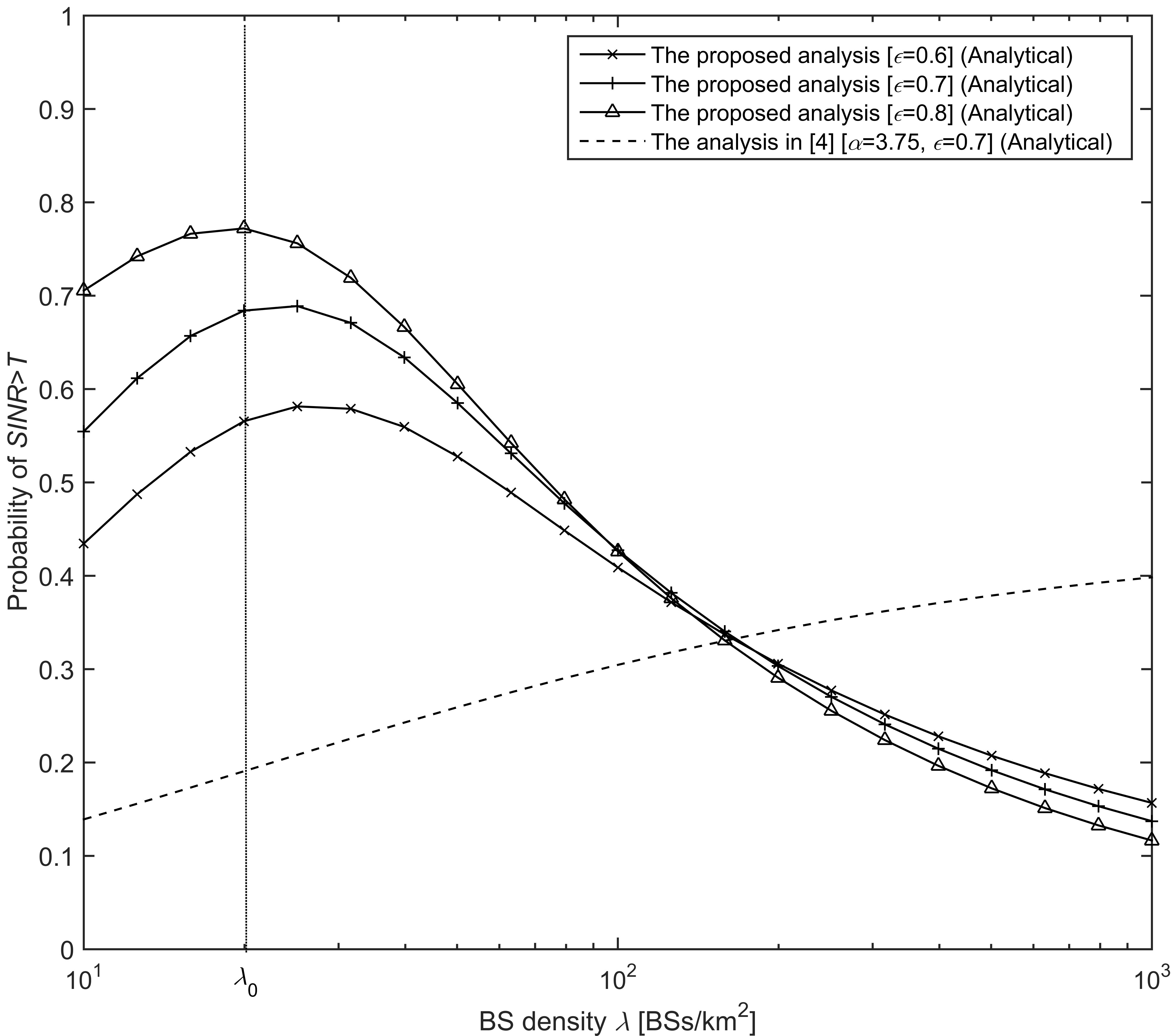}\renewcommand{\figurename}{Fig.}\caption{\label{fig:Coverage vs density}The coverage probability $P^{\mathrm{cov}}\left(\lambda,T\right)$
vs. the BS density with different $\epsilon$ and SINR threshold $T=0$
dB.}
\end{figure}

The results of $P^{\mathrm{cov}}\left(\lambda,T\right)$ against the
BS density for $T=0$ dB are plotted in Fig.~\ref{fig:Coverage vs density}.
From Fig.~\ref{fig:Coverage vs density}, we can observe that %
when considering both LoS and NLoS transmissions, the coverage probability
presents a significantly different behavior. When the SCN is sparse
and thus noise-limited, the coverage probability given by the proposed
analysis grows as $\lambda$ increases, similarly as that observed
in~\cite{Jeff_UL}. However, when the network is dense enough, the
coverage probability decreases as $\lambda$ increases, due to the
transition of a large number of interference paths from NLoS to LoS,
which is not captured in~\cite{Jeff_UL}. Particularly, during this
region, interference increases at a faster rate than the signal due
to the transition from mostly NLoS interference to LoS interference,
thereby causing a drop in the SINR hence the coverage probability.
In more detail, the coverage probability given by the proposed analysis
peaks at a certain density $\lambda_{0}$. When $\lambda$ increases
above $\lambda_{0}$, interfering UEs become closer to the typical
BS and their interfering signals start reaching the typical BS via
strong LoS paths. When $\lambda$ is further increased far above $\lambda_{0}$,
the coverage probability decreases at a slower pace because both the
signal power and the interference power are LoS dominated and increase
at approximately the same rate. There are still more and more interferers
whose signal reach the typical BS via LoS paths but their effect is
smaller than the dominating interferers.

It should also be noted that the coverage probability with different
FPC factor $\epsilon$ exhibits different trends. Specifically, when
the SCN is sparse, adopting a higher $\epsilon$ (e.g., $\epsilon=0.8$)
leads to a higher coverage probability. This is because the sparse
SCN is noise-limited and hence increasing the transmission power provides
better coverage performance. However, when the SCN is dense, adopting
a lower $\epsilon$ (e.g., $\epsilon=0.6$) leads to higher coverage
probability. This is because the dense SCN is interference-limited,
and the network experiences a surplus of strong LoS interference instead
of shortage of UL transmission power, and hence decreasing the transmission
power provides better coverage performance. Therefore, our results
suggest that in dense SCNs, increasing the UL transmission power may
degrade the coverage probability. Such observation is further investigated
in terms of ASE in the following subsection.

\subsection{The Results of $A^{\mathrm{ASE}}\left(\lambda,T_{0}\right)$ vs.
$\lambda$\label{subsec:ASE_vs_density}}

\begin{figure}
\centering{}\includegraphics[width=12cm]{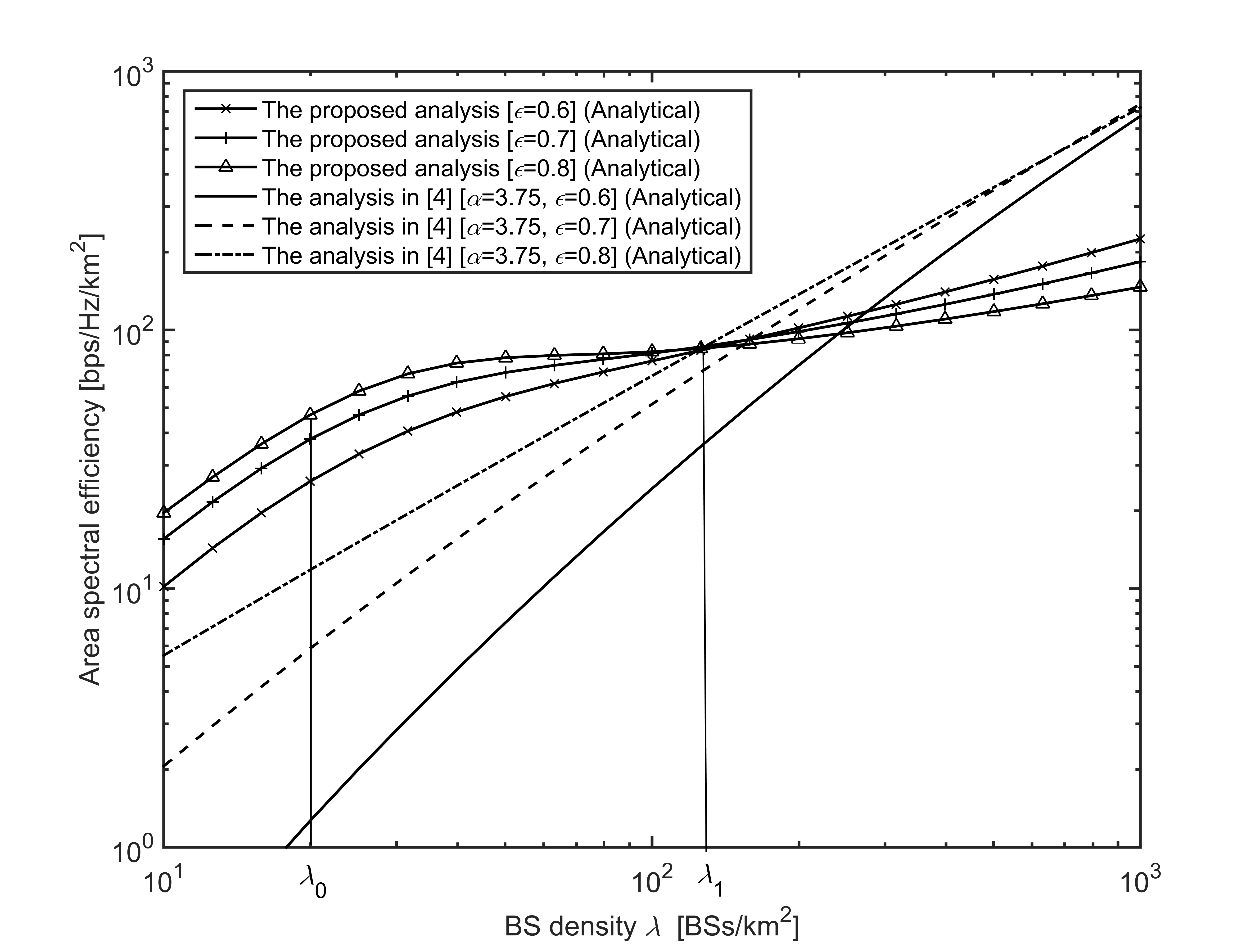}\renewcommand{\figurename}{Fig.}\caption{\label{fig:ASE}Area spectral efficiency $A^{\mathrm{ASE}}\left(\lambda,T_{0}\right)$
vs. the BS density with different $\epsilon$ and SINR threshold $T_{0}=0$
dB. $\lambda_{0}$ and $\lambda_{1}$ correspond to the BS density
when the ASE given by the proposed analysis starts to suffer from
a slow growth and when it starts to pick up the growth, respectively.}
\end{figure}

In this subsection, we investigate the ASE with $T_{0}$ = 0 dB based
on the analytical results of $P^{\textrm{cov}}\left(\lambda,T\right)$.
The results of $A^{\mathrm{ASE}}\left(\lambda,T_{0}\right)$ obtained
by comparing the proposed analysis with the analysis from~\cite{Jeff_UL}
are plotted in Fig.~\ref{fig:ASE}.

As can be seen from Fig.~\ref{fig:ASE}, the analysis from~\cite{Jeff_UL}
indicates that when the SCN is dense enough, the ASE increases linearly
with $\lambda$. In contrast, our proposed analysis reveals a more
complicated ASE trend. Specifically, when the SCN is relatively sparse,
i.e., $10^{0}\sim10^{1}\,\textrm{BSs/km}^{2}$, the ASE quickly increases
with $\lambda$ since the network is generally noise-limited, and
thus having UEs closer to their serving BSs improves performance.
When the SCN is extremely dense, i.e., around $10^{3}\,\textrm{BSs/km}^{2}$,
the ASE increases linearly with $\lambda$ because both the signal
power and the interference power are LoS dominated. As for the practical
range of $\lambda$ for the existing and the future cellular networks,
i.e., $10^{1}\sim10^{3}\,\textrm{BSs/km}^{2}$~\cite{Tutor_smallcell},
the ASE trend is interesting. First, when $\lambda\in\left[\lambda_{0},\lambda_{1}\right]$,
where $\lambda_{0}$ is around 20 and $\lambda_{1}$ ($\lambda_{1}>\lambda_{0}$)
is around 125 in Fig.~\ref{fig:ASE}, the ASE exhibits a slow-down
in the rate of growth %
due to the fast decrease of coverage probability shown in Fig.~\ref{fig:Coverage vs density}.
Thereafter, when $\lambda\geq\lambda_{1}$, the ASE exhibits an acceleration
in the growth rate due to the slow-down in the decrease of coverage
probability also shown in Fig.~\ref{fig:Coverage vs density}. Our
finding, the ASE may exhibits a slow-down in the rate of growth as
the BS density increases, is similar to our results reported for the
DL of SCNs~\cite{Ming_DL}, which indicates the significant impact
of the path loss model incorporating both NLoS and LoS transmissions.
Such impact makes a difference for dense SCNs in terms of the ASE
both quantitatively and qualitatively, comparing to that with a simplistic
path loss model that does not differentiate LoS and NLoS transmissions. 

Our proposed analysis also shows another important finding. A smaller
UL power compensation factor $\epsilon$ (e.g., $\epsilon=0.6$) can
greatly boost the ASE performance in 5G dense SCNs~\cite{Tutor_smallcell},
i.e., $10^{2}\sim10^{3}\,\textrm{BSs/km}^{2}$, while a larger $\epsilon$
(e.g., $\epsilon=0.8$) is more suitable for sparse SCNs, i.e., $10^{1}\sim10^{2}\,\textrm{BSs/km}^{2}$.
This contradicts the results in~\cite{Jeff_UL} where a larger UL
power compensation factor was predicted to always result in a better
ASE in the practical range of BS density, i.e., $10^{1}\sim10^{3}\,\textrm{BSs/km}^{2}$,
as shown in Fig.~\ref{fig:ASE}. Therefore, our theoretical analysis
indicates that the performance impact of LoS and NLoS transmissions
on UL SCNs with UL power compensation is also significant both quantitatively
and qualitatively, compared with the previous work in~\cite{Jeff_UL}
that does not differentiate LoS and NLoS transmissions. Interestingly,
our new finding implies that its is possible to save UE battery and
meanwhile achieve a high ASE in the UL of 5G dense SCNs, if $\epsilon$
is optimized. The intuition is that in dense SCNs, the network experiences
a surplus of strong LoS interference instead of shortage of UL transmission
power, and thus reducing the transmission powers of a large number
of interferers turns out to be a good strategy that enhances the ASE.
Note that our conclusion is made from the investigated set of parameters,
and it is of significant interest to further study the generality
of this conclusion in other network models and with other parameter
sets.

\subsection{Discussion on Various Values of $\alpha^{{\rm {L}}}$}

\begin{figure}
\centering{}\includegraphics[width=12cm]{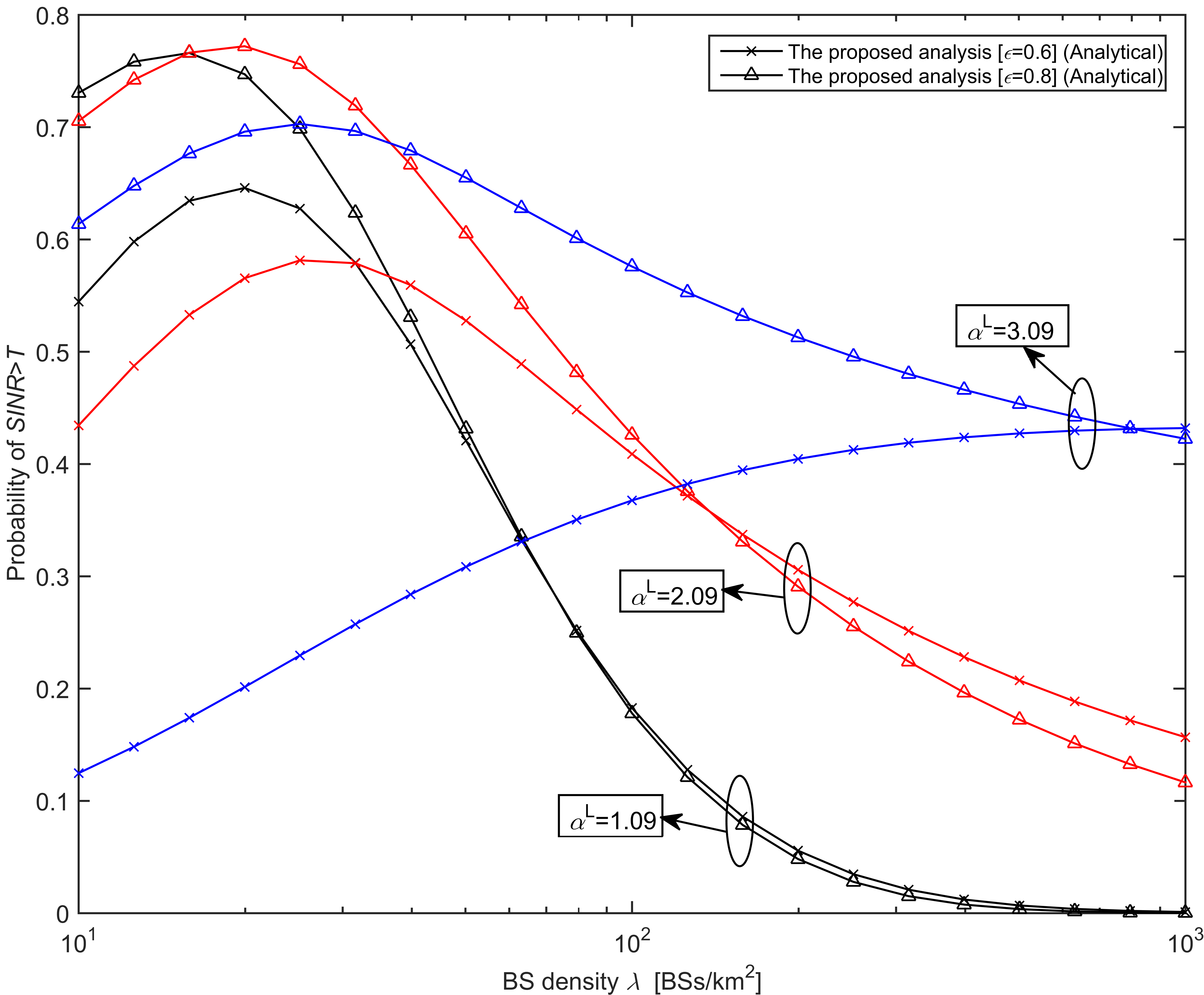}\renewcommand{\figurename}{Fig.}\caption{\label{fig:Pcov_Different_AlphaL}The coverage probability $P^{\mathrm{cov}}\left(\lambda,T\right)$
vs. the BS density with different $\epsilon$ and $\alpha^{\textrm{L}}$.
SINR threshold $T=0$ dB.}
\end{figure}

In this subsection, we change the value of $\alpha^{\textrm{L}}$
from 2.09 to 1.09 and 3.09, respectively, to investigate the performance
impact of $\alpha^{\textrm{L}}$. In Fig. \ref{fig:Pcov_Different_AlphaL},
the analytical results of $P^{\textrm{cov}}\left(\lambda,0\right)$
with $T_{0}$ = 0 dB and with various $\alpha^{\textrm{L}}$ and various
$\epsilon$ are compared.

As can be seen from Fig. \ref{fig:Pcov_Different_AlphaL}, the smaller
the $\alpha^{\textrm{L}}$, the larger the difference between the
NLoS path loss exponent $\alpha^{\textrm{NL}}$ and $\alpha^{\textrm{L}}$.
As a result, performance impact of the transition of interference
from the NLoS transmission to the LoS transmission becomes more drastic
as $\lambda$ increases. In other words, the slow growth %
of the $P^{\textrm{cov}}\left(\lambda,0\right)$ is more obvious to
observe. For example, when $\alpha^{\textrm{L}}$ takes a near-field
path loss exponent such as 1.09, the decrease of the $P^{\textrm{cov}}\left(\lambda,0\right)$
at $\lambda\in\left[\lambda_{0},\lambda_{1}\right]\textrm{ BSs/km}^{2}$
is substantial and it hardly recovers after $\lambda_{1}$.

As has been discussed in the subsection \ref{subsec:Coverage_vs_density},
when the SCN is sparse, adopting a higher $\epsilon$ leads to a higher
coverage probability. However, as $\lambda$ increases, adopting a
lower $\epsilon$ leads to a higher coverage probability. The BS density
around which the coverage probability with smaller $\epsilon$ surpasses
that with larger $\epsilon$ is defined as the transition point of
$\epsilon$. As can be seen from Fig. \ref{fig:Pcov_Different_AlphaL},
the transition point of various $\epsilon$ increases as $\alpha^{\textrm{L}}$
increases. It indicates that in dense SCNs with smaller $\alpha^{\textrm{L}}$,
the coverage probability using a smaller $\epsilon$ can soon outperform
that using a larger $\epsilon$ as the SCN becomes denser.

\subsection{Investigation of a Different Path Loss Model}

\begin{figure}
\centering{}\includegraphics[width=12cm]{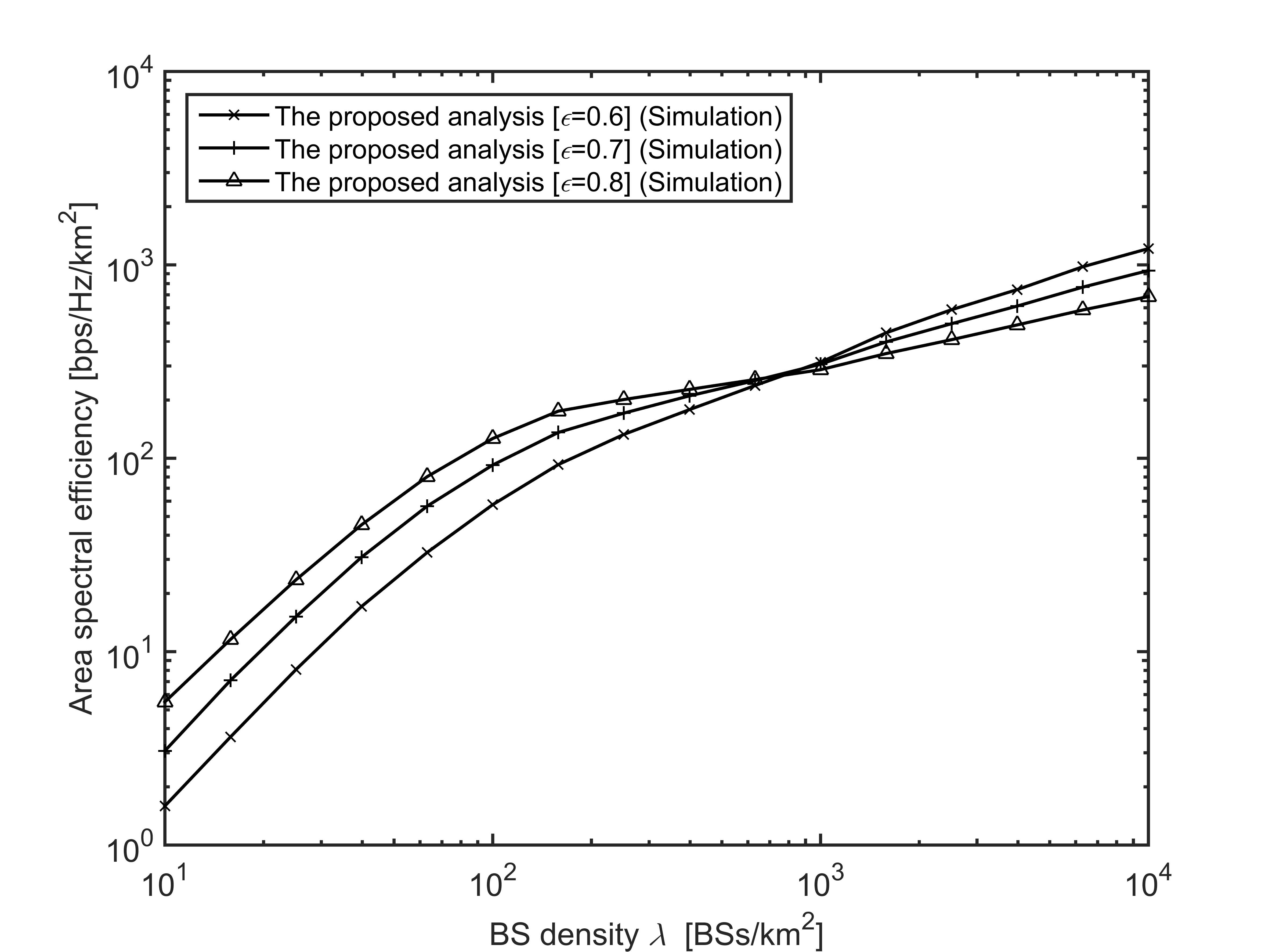}\renewcommand{\figurename}{Fig.}\caption{\label{fig:ASE_36828}Area spectral efficiency $A^{\mathrm{ASE}}\left(\lambda,T_{0}\right)$
vs. the BS density with the exponential LoS probability model, different
$\epsilon$ and SINR threshold $T_{0}=0$ dB.}
\end{figure}

In this subsection, we investigate the UL ASE performance assuming
a more complicated path loss model, in which the LoS probability is
defined as follows~\cite{TR36.828}

\begin{equation}
\textrm{P}\textrm{r}^{\textrm{L}}\left(r\right)=\begin{cases}
1-5\exp\left(-\frac{R_{1}}{r}\right), & 0<r\leq d_{1}\\
5\exp\left(-\frac{r}{R_{2}}\right), & r>d_{1}
\end{cases},\label{eq:Pr_LoS_exponential}
\end{equation}

\noindent where $R_{1}=0.156\textrm{ km}$, $R_{2}=0.03\textrm{ km}$,
and $d_{1}=\frac{R_{1}}{\ln10}$. The simulation results of the area
spectral efficiency $A^{\mathrm{ASE}}\left(\lambda,T_{0}\right)$
vs. the BS density is shown in Fig. \ref{fig:ASE_36828}.

As can be seen from Fig. \ref{fig:ASE_36828}, the area spectral efficiency
with the exponential LoS probability model exhibits a slow-down in
the rate of growth in certain BS density regions, which qualitatively
confirms our observations in subsection \ref{subsec:ASE_vs_density}
with the linear LoS probability model. Specifically, in Fig. \ref{fig:ASE_36828},
the numerical result for $\lambda_{0}$ is around $10^{2}\,\textrm{BSs/km}^{2}$.
Furthermore, the area spectral efficiency with the exponential LoS
probability model exhibits a similar trend as discussed in subsection
\ref{subsec:ASE_vs_density} with the linear LoS probability model,
i.e., using a smaller UL power compensation factor $\epsilon$ can
outperform that using a larger $\epsilon$ as the SCN becomes denser.

\subsection{Investigation of the Performance Impact of Ricean Fading}

\begin{figure}
\noindent \begin{centering}
\textbf{\includegraphics[width=10cm]{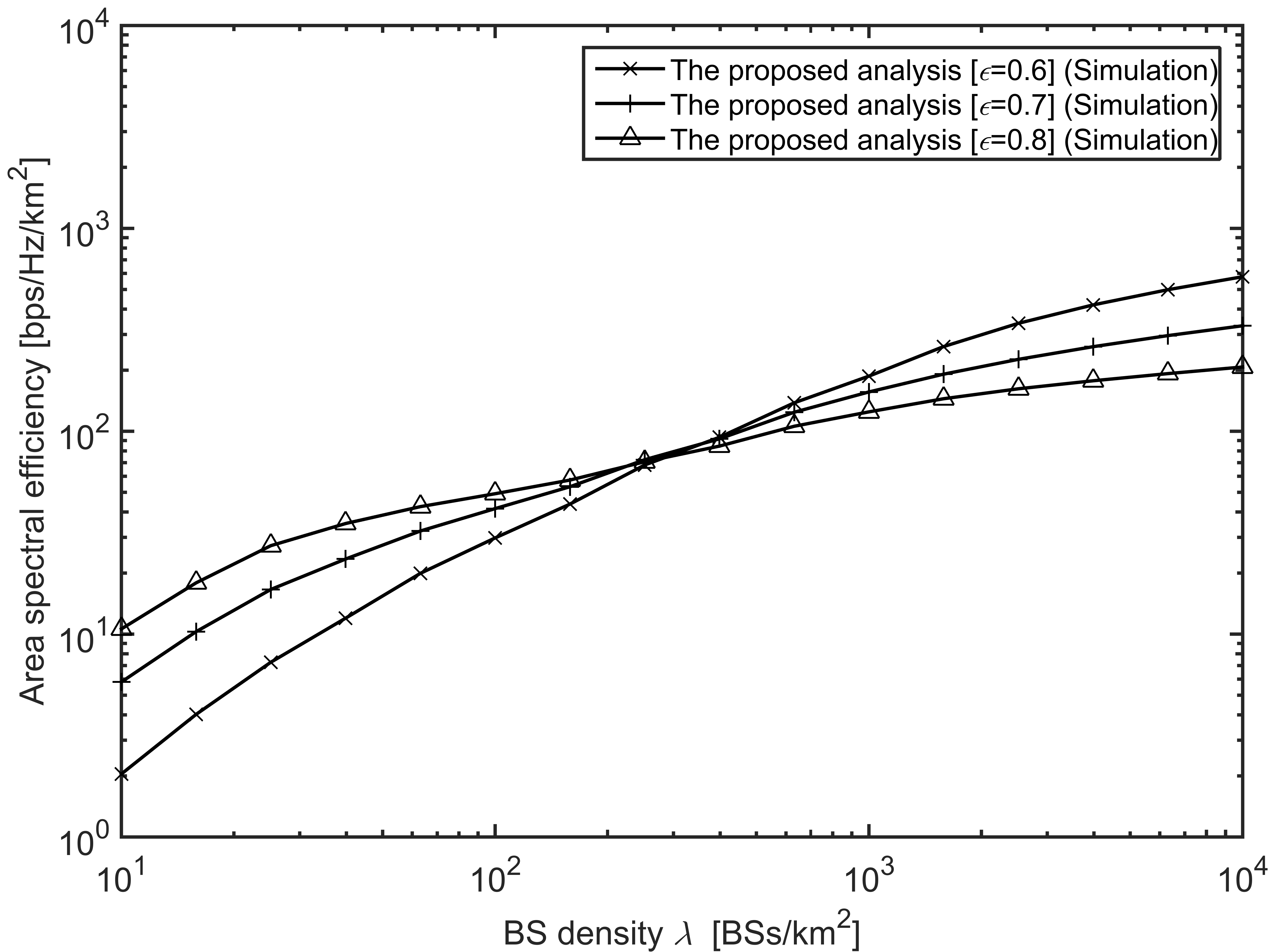}}\renewcommand{\figurename}{Fig.}
\par\end{centering}
\noindent \centering{}\textbf{\caption{\textbf{\label{fig:ASE_Ricean_Fading}Area spectral efficiency $A^{\mathrm{ASE}}\left(\lambda,T_{0}\right)$
vs. the BS density with the linear LoS probability model, different
$\epsilon$ and SINR threshold $T_{0}=0$ dB, including the Ricean
fading. }}
}
\end{figure}

\textbf{In this subsection, we investigate the UL ASE performance
assuming a linear path loss model including the Ricean fading. Here
we adopt a practical model of Ricean fading}~\cite{Jesus_LoS_DL_ICC16}\textbf{
with $K$ factor $K=15$ dB. The simulation results of the area spectral
efficiency $A^{\mathrm{ASE}}\left(\lambda,T_{0}\right)$ vs. the BS
density is shown in Fig. \ref{fig:ASE_Ricean_Fading}.}

\textbf{}

\textbf{As can be seen from Fig. \ref{fig:ASE_Ricean_Fading}, the
area spectral efficiency with the linear LoS probability model and
the Ricean fading exhibits a slow-down in the rate of growth as the
BS density increases, which qualitatively confirms our observations
in subsection \ref{subsec:ASE_vs_density} for the linear LoS probability
model and the Rayleigh fading. Furthermore, the area spectral efficiency
with the Ricean fading exhibits a similar trend as discussed in subsection
\ref{subsec:ASE_vs_density} with the Rayleigh fading, i.e., using
a smaller UL power compensation factor $\epsilon$ can outperform
that using a larger $\epsilon$ as the SCN becomes denser. Since the
simulation results of Ricean fading and Rayleigh fading are not qualitatively
different, we suggest to use a simplified model with the Rayleigh
fading in theoretical analysis.}

\section{Conclusion\label{sec:Conclusion}}

In this paper, we have investigated the impact of a piecewise linear
path loss model incorporating both LoS and NLoS transmissions in the
performance of the UL of dense SCNs. Analytical results were obtained
for the coverage probability and the ASE performance. The results
show that LoS and NLoS transmissions have a significant impact in
the ASE of the UL of dense SCNs, both quantitatively and qualitatively,
compared with previous works that does not differentiate LoS and NLoS
transmissions. Specifically, we found that 
\begin{itemize}
\item The ASE may suffer from a slow growth as the UE density increases
in the UL of dense SCNs.
\item The ASE with a smaller UL power compensation factor considerably outperforms
that with a larger UL power compensation factor in dense SCNs. The
reverse is true for sparse SCNs.
\end{itemize}
As our future work, we will consider other factors of realistic networks
in the theoretical analysis for SCNs, such as the introduction of
Ricean fading or Nakagami fading, because the multi-path fading model
is also affected by the LoS and NLoS transmissions.

\section*{Appendix A: Proof of Theorem \ref{thm:p_cov_UL}}

\begin{doublespace}
Given the piecewise path loss model presented in Section~\ref{sec:Network-Model},
$P^{\textrm{cov}}\left(\lambda,T\right)$ can be derived as

\vspace*{-0.5cm}

\begin{equation}
\begin{array}{l}
\hspace{-0.1cm}\hspace{-0.1cm}P^{\mathrm{cov}}\left(\lambda,T\right)\\
=\int_{0}^{\infty}\textrm{Pr}\left[\left.\textrm{SINR}>T\right|r\right]f_{R}\left(r\right)dr\\
=\int_{0}^{\infty}\textrm{Pr}\left[\frac{P_{0}g\zeta\left(r\right)^{\left(\epsilon-1\right)}}{\sigma^{2}+I_{Z}}>T\right]f_{R}\left(r\right)dr\\
=\int_{0}^{d_{1}}\textrm{Pr}\left[\left.\frac{P_{0}g\left(A^{\textrm{L}}r^{\alpha^{\mathrm{\textrm{L}}}}\right)^{\left(\epsilon-1\right)}}{\sigma^{2}+I_{Z}}>T\right|\mathrm{LoS}\right]f_{R,1}^{\textrm{L}}\left(r\right)\hspace{-0.1cm}dr\\
+\int_{0}^{d_{1}}\textrm{Pr}\left[\left.\frac{P_{0}g\left(A^{\textrm{NL}}r^{\alpha^{\mathrm{\textrm{NL}}}}\right)^{\left(\epsilon-1\right)}}{\sigma^{2}+I_{Z}}>T\right|\hspace{-0.1cm}\mathrm{NLoS}\right]f_{R,1}^{\textrm{NL}}\left(r\right)dr\\
+\ldots\\
+\int_{d_{N-1}}^{\infty}\textrm{Pr}\left[\left.\frac{P_{0}g\left(A^{\textrm{L}}r^{\alpha^{\mathrm{\textrm{L}}}}\right)^{\left(\epsilon-1\right)}}{\sigma^{2}+I_{Z}}>T\right|\hspace{-0.1cm}\mathrm{LoS}\right]f_{R,N}^{\textrm{L}}\left(r\right)\hspace{-0.1cm}dr\\
+\int_{d_{N-1}}^{\infty}\textrm{Pr}\left[\left.\frac{P_{0}g\left(A^{\textrm{NL}}r^{\alpha^{\mathrm{\textrm{NL}}}}\right)^{\left(\epsilon-1\right)}}{\sigma^{2}+I_{Z}}>T\right|\hspace{-0.1cm}\mathrm{NLoS}\right]f_{R,N}^{\textrm{NL}}\left(r\right)dr\\
\triangleq\overset{N}{\underset{n=1}{\sum}}\left(T_{n}^{\mathrm{L}}+T_{n}^{\mathrm{NL}}\right).
\end{array}\label{eq:proof_of_Pcov}
\end{equation}

In the following, we show how to compute $f_{R,n}^{\textrm{L}}\left(r\right)$
and $f_{R,n}^{\textrm{NL}}\left(r\right)$.
\end{doublespace}

To compute $f_{R,n}^{\textrm{L}}\left(r\right)$, we define two events
as follows

\begin{doublespace}
Event $B^{\textrm{L}}$: The nearest BS with a LoS path to the UE
is located at distance $X^{\textrm{L}}$. The CCDF of $X^{\textrm{L}}$
is written as $\bar{F}_{X}^{\textrm{L}}\left(x\right)=\exp\left(-\int_{0}^{x}\Pr^{\textrm{L}}\left(u\right)2\pi u\lambda du\right)$~\cite{Ming_DL}.
Taking the derivative of $\left(1-\bar{F}_{X}^{\textrm{L}}\left(x\right)\right)$
with regard to $x$, we can get the PDF of $X^{\textrm{L}}$ as
\end{doublespace}

\vspace*{-0.5cm}

\begin{equation}
f_{X}^{\textrm{L}}\left(x\right)=\exp\left(-\int_{0}^{x}\textrm{Pr}^{\textrm{L}}\left(u\right)2\pi u\lambda du\right)\Pr^{\textrm{L}}\left(x\right)2\pi x\lambda.\label{eq:PDF_X_L}
\end{equation}

\begin{doublespace}
Event $C^{\textrm{NL}}$ conditioned on the value of $X^{\textrm{L}}$:
Given that $X^{\textrm{L}}=x$, the nearest BS with a NLoS path to
the UE is located farther than distance $x_{1}$, where $A^{\textrm{L}}x^{\alpha^{\textrm{L}}}=A^{\textrm{NL}}x_{1}^{\alpha^{\textrm{NL}}}$,
and $x_{1}=\left(A^{\textrm{L}}x^{\alpha^{\textrm{L}}}/A^{\textrm{NL}}\right)^{1/\alpha^{\textrm{NL}}}$.
The conditional probability of $C^{\textrm{NL}}$ on condition of
$X^{\textrm{L}}=x$ can be computed by
\end{doublespace}

\vspace*{-0.5cm}

\begin{equation}
\Pr\left[\left.C^{\textrm{NL}}\right|X^{\textrm{L}}=x\right]=\exp\left(-\int_{0}^{x_{1}}\left(1-\mathrm{Pr^{L}}\left(u\right)\right)2\pi u\lambda du\right).\label{eq:Pr_CNL}
\end{equation}

Then, we consider the event that the UE is associated with a BS with
a LoS path and such BS is located at distance $R_{n}^{\textrm{L}}$.
$f_{R,n}^{\textrm{L}}\left(r\right)$ can be derived as

\begin{equation}
\begin{array}{l}
f_{R,n}^{\textrm{L}}\left(r\right)\\
=f_{X}^{\textrm{L}}\left(r\right)\Pr\left[\left.C^{\textrm{NL}}\right|X^{\textrm{L}}=r\right]\\
=\exp\left(-\int_{0}^{r}\Pr^{\textrm{L}}\left(u\right)2\pi u\lambda du\right)\Pr^{\textrm{L}}\left(r\right)2\pi r\lambda\\
\times\exp\left(-\int_{0}^{r_{1}}\left(1-\Pr^{\textrm{L}}\left(u\right)\right)2\pi u\lambda du\right),~~\left(d_{n-1}<r<d_{n}\right).
\end{array}\label{eq:PDF_RL}
\end{equation}

Having obtained $f_{R,n}^{\textrm{L}}\left(r\right)$, we move on
to evaluate $\mathrm{\textrm{Pr}}\left[\left.\frac{P_{0}g\left(A^{\textrm{L}}r^{\alpha^{\mathrm{\textrm{L}}}}\right)^{\left(\epsilon-1\right)}}{\sigma^{2}+I_{Z}}>T\right|\textrm{LoS}\right]$
in (\ref{eq:Pcov_LoS}) as

\begin{equation}
\begin{array}{l}
\mathrm{\textrm{Pr}}\left[\left.\frac{P_{0}g\left(A^{\textrm{L}}r^{\alpha^{\mathrm{\textrm{L}}}}\right)^{\left(\epsilon-1\right)}}{\sigma^{2}+I_{Z}}>T\right|\textrm{LoS}\right]\\
=\mathrm{\textrm{Pr}}\left[\left.g>\frac{T\left(\sigma^{2}+I_{Z}\right)}{P_{0}\left(A^{\textrm{L}}r^{\alpha^{\mathrm{\textrm{L}}}}\right)^{\left(\epsilon-1\right)}}\right|\textrm{LoS}\right]\\
=\mathbb{E}_{I_{Z}}\left\{ \exp\left(-\frac{T\left(\sigma^{2}+I_{Z}\right)}{P_{0}\left(A^{\textrm{L}}r^{\alpha^{\textrm{L}}}\right)^{\left(\epsilon-1\right)}}\right)\right\} \\
=\exp\left(-\frac{T\sigma^{2}}{P_{0}\left(A^{\textrm{L}}r^{\alpha^{\textrm{L}}}\right)^{\left(\epsilon-1\right)}}\right)\mathbb{E}_{I_{Z}}\left\{ \exp\left(-\frac{TI_{Z}}{P_{0}\left(A^{\textrm{L}}r^{\alpha^{\textrm{L}}}\right)^{\left(\epsilon-1\right)}}\right)\right\} \\
=\exp\left(-\frac{T\sigma^{2}}{P_{0}\left(A^{\textrm{L}}r^{\alpha^{\textrm{L}}}\right)^{\left(\epsilon-1\right)}}\right)\mathscr{L}_{I_{Z}}\left(\frac{T}{P_{0}\left(A^{\textrm{L}}r^{\alpha^{\textrm{L}}}\right)^{(\epsilon-1)}}\right),
\end{array}\label{eq:Derivation_Pcov_LoS}
\end{equation}

\noindent where $\mathscr{L}_{I_{Z}}\left(s\right)$ is the Laplace
transform of RV $I_{Z}$ evaluated at $s$.

To compute $f_{R,n}^{\textrm{NL}}\left(r\right)$, we define two events
as follows

\begin{doublespace}
Event $B^{\textrm{NL}}$: The nearest BS with a NLoS path to the UE
is located at distance $X^{\textrm{NL}}$. The CCDF of $X^{\textrm{NL}}$
is written as $\bar{F}_{X}^{\textrm{NL}}\left(x\right)=\exp\left(-\int_{0}^{x}\left(1-\Pr^{\textrm{L}}\left(u\right)\right)2\pi u\lambda du\right)$.
Taking the derivative of $\left(1-\bar{F}_{X}^{\textrm{L}}\left(x\right)\right)$
with regard to $x$, we can get the PDF of $X^{\textrm{NL}}$ as
\end{doublespace}

\begin{equation}
f_{X}^{\textrm{NL}}\left(x\right)=\exp\left(-\int_{0}^{x}\left(1-\mathrm{Pr^{L}}\left(u\right)\right)2\pi u\lambda du\right)\left(1-\mathrm{Pr^{L}}\left(x\right)\right)2\pi x\lambda.\label{eq:PDF_XNL}
\end{equation}

\begin{doublespace}
Event $C^{\textrm{L}}$ conditioned on the value of $X^{\textrm{NL}}$:
Given that $X^{\textrm{NL}}=x$, the nearest BS with a LoS path to
the UE is located farther than distance $x_{2}$, where $A^{\textrm{L}}x_{2}^{\alpha^{\textrm{L}}}=A^{\textrm{NL}}x^{\alpha^{\textrm{NL}}}$,
and $x_{2}=\left(A^{\textrm{NL}}x^{\alpha^{\textrm{NL}}}/A^{\textrm{L}}\right)^{1/\alpha^{\textrm{L}}}$.
The conditional probability of $C^{\textrm{L}}$ on condition of $X^{\textrm{NL}}=x$
can be computed by
\end{doublespace}

\begin{equation}
\Pr\left[\left.C^{\textrm{L}}\right|X^{\textrm{NL}}=x\right]=\begin{cases}
\exp\left(-\int_{0}^{x_{2}}\left(\mathrm{Pr^{L}}\left(u\right)\right)2\pi u\lambda du\right), & 0<x\leq y_{1}\\
\exp\left(-\int_{0}^{d_{1}}\left(\mathrm{Pr^{L}}\left(u\right)\right)2\pi u\lambda du\right), & x>y_{1}
\end{cases}.\label{eq:Pr_CL}
\end{equation}

Then, we consider the event that the UE is associated with a BS with
a NLoS path and such BS is located at distance $R_{n}^{\textrm{NL}}$.
$f_{R,n}^{\textrm{NL}}\left(r\right)$ can be derived as

\begin{equation}
\begin{array}{l}
f_{R,n}^{\textrm{NL}}\left(r\right)\\
=f_{X}^{\textrm{NL}}\left(r\right)\Pr\left[\left.C^{\textrm{L}}\right|X^{\textrm{NL}}=r\right]\\
=\exp\left(-\int_{0}^{r}\left(1-\Pr^{\textrm{L}}\left(u\right)\right)2\pi u\lambda du\right)\left(1-\Pr^{\textrm{L}}\left(r\right)\right)2\pi r\lambda\\
\times\exp\left(-\int_{0}^{r_{2}}\left(\Pr^{\textrm{L}}\left(u\right)\right)2\pi u\lambda du\right),~~\left(d_{n-1}<r<d_{n}\right).
\end{array}\label{eq:PDF_RNL}
\end{equation}

Similar to (\ref{eq:Derivation_Pcov_LoS}), $\textrm{Pr}\left[\left.\frac{P_{0}g\left(A^{\textrm{NL}}r^{\alpha^{\textrm{NL}}}\right)^{\left(\epsilon-1\right)}}{\sigma^{2}+I_{Z}}>T\right|\textrm{NLoS}\right]$
can be computed by

\begin{equation}
\begin{array}{l}
\textrm{Pr}\left[\left.\frac{P_{0}g\left(A^{\textrm{NL}}r^{\alpha^{\textrm{NL}}}\right)^{\left(\epsilon-1\right)}}{\sigma^{2}+I_{Z}}>T\right|\textrm{NLoS}\right]\\
=\mathbb{E}_{I_{Z}}\left\{ \exp\left(-\frac{T\left(\sigma^{2}+I_{Z}\right)}{P_{0}\left(A^{\textrm{NL}}r^{\alpha^{\textrm{NL}}}\right)^{\left(\epsilon-1\right)}}\right)\right\} \\
=\exp\left(-\frac{T\sigma^{2}}{P_{0}\left(A^{\textrm{NL}}r^{\alpha^{\textrm{NL}}}\right)^{\left(\epsilon-1\right)}}\right)\mathscr{L}_{I_{Z}}\left(\frac{T}{P_{0}\left(A^{\textrm{NL}}r^{\alpha^{\textrm{NL}}}\right)^{(\epsilon-1)}}\right).
\end{array}\label{eq:Derivation_Pcov_NLoS}
\end{equation}

\begin{doublespace}
Our proof is completed by applying the definition of $T_{n}^{\mathrm{L}}$
and $T_{n}^{\mathrm{NL}}$ in (\ref{eq:Pcov_thm1}).
\end{doublespace}

\section*{Appendix B: Proof of Lemma \ref{lem:T1L}}

\begin{doublespace}
Based on (\ref{eq:Pcov_case1}), $T_{1}^{\textrm{L}}$ can be obtained
as

\begin{equation}
\begin{array}{l}
T_{1}^{\textrm{L}}\\
=\int_{0}^{d_{1}}\mathrm{\textrm{Pr}}\left[\left.\frac{P_{0}g\left(A^{\textrm{L}}r^{\alpha^{\mathrm{\textrm{L}}}}\right)^{\left(\epsilon-1\right)}}{\sigma^{2}+I_{Z}}>T\right|\textrm{LoS}\right]f_{R,1}^{\textrm{L}}\left(r\right)\hspace{-0.1cm}dr\\
=\int_{0}^{d_{1}}\exp\left(-\frac{T\sigma^{2}}{P_{0}\left(A^{L}r^{\alpha^{\textrm{L}}}\right)^{\left(\epsilon-1\right)}}\right)\mathscr{L}_{I_{Z}}\left(\frac{T}{P_{0}\left(A^{L}r^{\alpha^{\textrm{L}}}\right)^{(\epsilon-1)}}\right)f_{R,1}^{\textrm{L}}\left(r\right)dr.
\end{array}\label{eq:Proof_T1L}
\end{equation}

The Laplace transform $\mathscr{L}_{I_{Z}}\left(s\right)$ is expressed
as

\begin{equation}
\begin{array}{l}
\mathscr{L}_{I_{Z}}\left(s\right)\\
=\mathbb{E}_{I_{Z}}\left[\exp\left(-sI_{Z}\right)\right]\\
=\mathbb{E}_{r_{z},d_{z},g_{z}}\left[\exp\left(-s\underset{Z}{\sum}P_{0}\beta\left(r_{z}\right)\zeta\left(d_{z}\right)^{-1}g_{z}\right)\right]\\
=\mathbb{E}_{r_{z},d_{z}}\left[\underset{Z}{\prod}\mathbb{E}_{g_{z}}\left(\exp\left(-sP_{0}\beta\left(r_{z}\right)\zeta\left(d_{z}\right)^{-1}g_{z}\right)\right)\right]\\
=\mathbb{E}_{r_{z},d_{z}}\left[\underset{Z}{\prod}\frac{1}{1+sP_{0}\beta\left(r_{z}\right)\zeta\left(d_{z}\right)^{-1}}\right]\\
=\exp\left(-2\pi\lambda\int_{r}^{\infty}\left(1-\mathbb{E}_{r_{z}}\left[\frac{1}{1+sP_{0}\beta\left(r_{z}\right)\zeta\left(x\right)^{-1}}\right]\right)xdx\right)\\
=\exp\left(-2\pi\lambda\int_{r}^{\infty}\mathbb{E}_{r_{z}}\left[\frac{1}{1+s^{-1}P_{0}^{-1}\beta\left(r_{z}\right)^{-1}\zeta\left(x\right)}\right]xdx\right)\\
=\exp\left(-2\pi\lambda\int_{r}^{d_{1}}\left(1-\frac{x}{d_{1}}\right)\mathbb{E}_{r_{z}}\left[\left.\frac{1}{1+s^{-1}P_{0}^{-1}\beta\left(r_{z}\right)^{-1}\zeta\left(x\right)}\right|\textrm{LoS}\right]xdx\right)\\
\times\exp\left(-2\pi\lambda\int_{r_{1}}^{d_{1}}\left(\frac{x}{d_{1}}\right)\mathbb{E}_{r_{z}}\left[\left.\frac{1}{1+s^{-1}P_{0}^{-1}\beta\left(r_{z}\right)^{-1}\zeta\left(x\right)}\right|\textrm{NLoS}\right]xdx\right)\\
\times\exp\left(-2\pi\lambda\int_{d_{1}}^{\infty}\mathbb{E}_{r_{z}}\left[\left.\frac{1}{1+s^{-1}P_{0}^{-1}\beta\left(r_{z}\right)^{-1}\zeta\left(x\right)}\right|\textrm{NLoS}\right]xdx\right),
\end{array}\label{eq:Laplace}
\end{equation}

\end{doublespace}

\begin{doublespace}
\noindent where the expectation function averaged over $r_{z}$ is
derived as follows
\end{doublespace}

\begin{doublespace}
\begin{equation}
\mathbb{E}_{r_{z}}\left[\frac{1}{1+s^{-1}P_{0}^{-1}\beta\left(r_{z}\right)^{-1}\zeta\left(x\right)}|\textrm{LoS}\right]=\int_{0}^{\infty}\left(\frac{1}{1+s^{-1}P_{0}^{-1}\beta\left(u\right)^{-1}\zeta\left(x\right)}f_{R_{z}}\left(u\right)du\right).\label{eq:ERz}
\end{equation}

\end{doublespace}

\begin{doublespace}
By plugging (\ref{eq:ERz}) into (\ref{eq:Laplace}), we can obtain
(\ref{eq:Laplace_transform}).
\end{doublespace}

\section*{Appendix C: Proof of Lemma \ref{lem:G-L quadrature}}

By using the change of variable $\pi\lambda r^{2}\rightarrow\widetilde{r}$,
(\ref{eq:T2NL}) can be rewritten as

\vspace{0.2cm}

\noindent 
\begin{equation}
T_{2}^{\textrm{NL}}\hspace{-0.1cm}=\hspace{-0.1cm}\int_{\pi\lambda d_{1}^{2}}^{\infty}\exp\left(-\frac{T\sigma^{2}}{P_{0}\left(\sqrt{\widetilde{r}\left(\pi\lambda\right)^{-1}}\right)^{\alpha^{\textrm{NL}}\left(\epsilon-1\right)}}\right)\mathscr{L}_{I_{Z}}\hspace{-0.1cm}\left(\hspace{-0.1cm}\frac{T}{P_{0}\left(\sqrt{\widetilde{r}\left(\pi\lambda\right)^{-1}}\right)^{\alpha^{\textrm{NL}}\left(\epsilon-1\right)}}\hspace{-0.1cm}\right)e^{-\widetilde{r}}d\widetilde{r}.\label{eq:variable_change_of_T2NL}
\end{equation}

By using the change of variable $\widetilde{r}-\pi\lambda\left(d_{1}\right)^{2}\rightarrow v$,
(\ref{eq:variable_change_of_T2NL}) can be rewritten as

\vspace{0.2cm}

\begin{equation}
\begin{array}{l}
T_{2}^{\textrm{NL}}=\int_{0}^{\infty}\exp\left(-\frac{T\sigma^{2}}{P_{0}\left(\sqrt{\left[v+\pi\lambda\left(d_{1}\right)^{2}\right]\left(\pi\lambda\right)^{-1}}\right)^{\alpha^{\textrm{NL}}\left(\epsilon-1\right)}}\right)\\
\times\hspace{-0.1cm}\mathscr{L}_{I_{Z}}\hspace{-0.1cm}\left(\hspace{-0.1cm}\frac{T}{P_{0}\left(\sqrt{\left[v+\pi\lambda\left(d_{1}\right)^{2}\right]\left(\pi\lambda\right)^{-1}}\right)^{\alpha^{\textrm{NL}}\left(\epsilon-1\right)}}\hspace{-0.1cm}\right)e^{-\pi\lambda\left(d_{1}\right)^{2}}e^{-v}dv.
\end{array}\label{eq:second_variable_change_of_T2NL}
\end{equation}

By using the method of Gauss-Laguerre quadrature as shown in (\ref{eq:Gauss_Laguerre_Quadrature}),
we complete the proof.

\end{document}